\documentclass[aps,prb,twocolumn,superscriptaddress,10pt]{revtex4-1}
\usepackage{graphicx} % Required for inserting images
\usepackage[utf8]{inputenc}
\usepackage[version=4]{mhchem}
\usepackage{caption}
\usepackage{subcaption}

\begin{document}

\title{Influence of Hydrogen-Incorporation on the Bulk Electronic Structure and Chemical Bonding in Palladium}

% Force line breaks with \\
\author{L.~J.~Bannenberg}
 \affiliation{Faculty of Applied Sciences, Delft University of Technology, Mekelweg 15, 2629 JB Delft, The Netherlands.}

\author{F.~Garc\'{i}a-Mart\'{i}nez}
\affiliation{Photon Science, Deutsches Elektronen-Synchrotron DESY, 22607 Hamburg, Germany.}

\author{P.~L\"{o}mker}
\affiliation{Department of Physics, Stockholm University, 10691 Stockholm, Sweden.}
\affiliation{Wallenberg Initiative Materials Science for Sustainability, Department of Physics, Stockholm University, 114 28 Stockholm, Sweden.}
\affiliation{Photon Science, Deutsches Elektronen-Synchrotron DESY, 22607 Hamburg, Germany.}

\author{R.~Y.~Engel}
 \affiliation{Department of Physics, Stockholm University, 10691 Stockholm, Sweden.}
 \affiliation{Wallenberg Initiative Materials Science for Sustainability, Department of Physics, Stockholm University, 114 28 Stockholm, Sweden.}

\author{C.~Schlueter}
\affiliation{Photon Science, Deutsches Elektronen-Synchrotron DESY, 22607 Hamburg, Germany.}

\author{H.~Schreuders}
\author{A.~Navarathna}
 \affiliation{Faculty of Applied Sciences, Delft University of Technology, Mekelweg 15, 2629 JB Delft, The Netherlands.}

\author{L.~E.~Ratcliff}%
\affiliation{Centre for Computational Chemistry, School of Chemistry, University of Bristol, Bristol BS8 1TS, United Kingdom.%\\
}%
\affiliation{Hylleraas Centre for Quantum Molecular Sciences, Department of Chemistry, UiT The Arctic University of Norway, N-9037 Troms\o{}, Norway.}
 
\author{A.~Regoutz}
  \email{anna.regoutz@chem.ox.ac.uk}
\affiliation{Department of Chemistry, University of Oxford, Inorganic Chemistry Laboratory, Oxford OX1 3QR, United Kingdom.}
 \affiliation{Department of Chemistry, University College London, London WC1H 0AJ, United Kingdom.}

\begin{abstract}
Palladium hydride is a model system for studying metal-hydrogen interactions. Yet, its bulk electronic structure has proven difficult to directly probe, with most studies to date limited to surface-sensitive photoelectron spectroscopy approaches. This work reports the first in-situ ambient-pressure hard X-ray photoelectron spectroscopy (AP-HAXPES) study of hydrogen incorporation in Pd thin films, providing direct access to bulk chemical and electronic information at elevated hydrogen pressures. Structural characterisation by in-situ X-ray diffraction and neutron reflectometry under comparable conditions establishes a direct correlation between hydrogen loading, lattice expansion, and electronic modifications. Comparison with density functional theory (DFT) reveals how hydrogen stoichiometry and site occupancy govern the density of occupied states near the Fermi level. These results resolve long-standing questions regarding PdH and establish AP-HAXPES as a powerful tool for probing the bulk electronic structure of metal hydrides under realistic conditions.
\end{abstract}

\maketitle

\section{Introduction}
Palladium hydride (PdH$_x$) is one of the archetype metal hydrides, with Pd being the first metal that was found to be capable of absorbing large quantities of hydrogen in the latter half of the 19th century.~\cite{graham1869} What makes Pd special among the metals forming hydrides is that it can catalyse the hydrogen dissociation reaction and absorb vast amounts of hydrogen at modest pressures at the same time. These properties ensure many of palladium hydride's applications in catalysis, hydrogen compressors, purification membranes, and hydrogen sensors, amongst others.~\cite{butler1991,adams2011,lototskyy2014,al2015,darmadi2020,bannenberg2020,zhang2022palladium} Moreover, PdH$_x$ has been a popular playground to study the inherent size- and shape-dependence of the thermodynamics of metal hydrides, which can be drastically different for Pd nanoparticles and thin films due to e.g.\ surface, stress/strain effects and interactions with the support.~\cite{yamauchi2008,baldi2014,Griessen2016,sytwu2018,suzana2021,zhou2023}

\begin{figure}[ht]
\centering
    \includegraphics[width=0.9\linewidth]{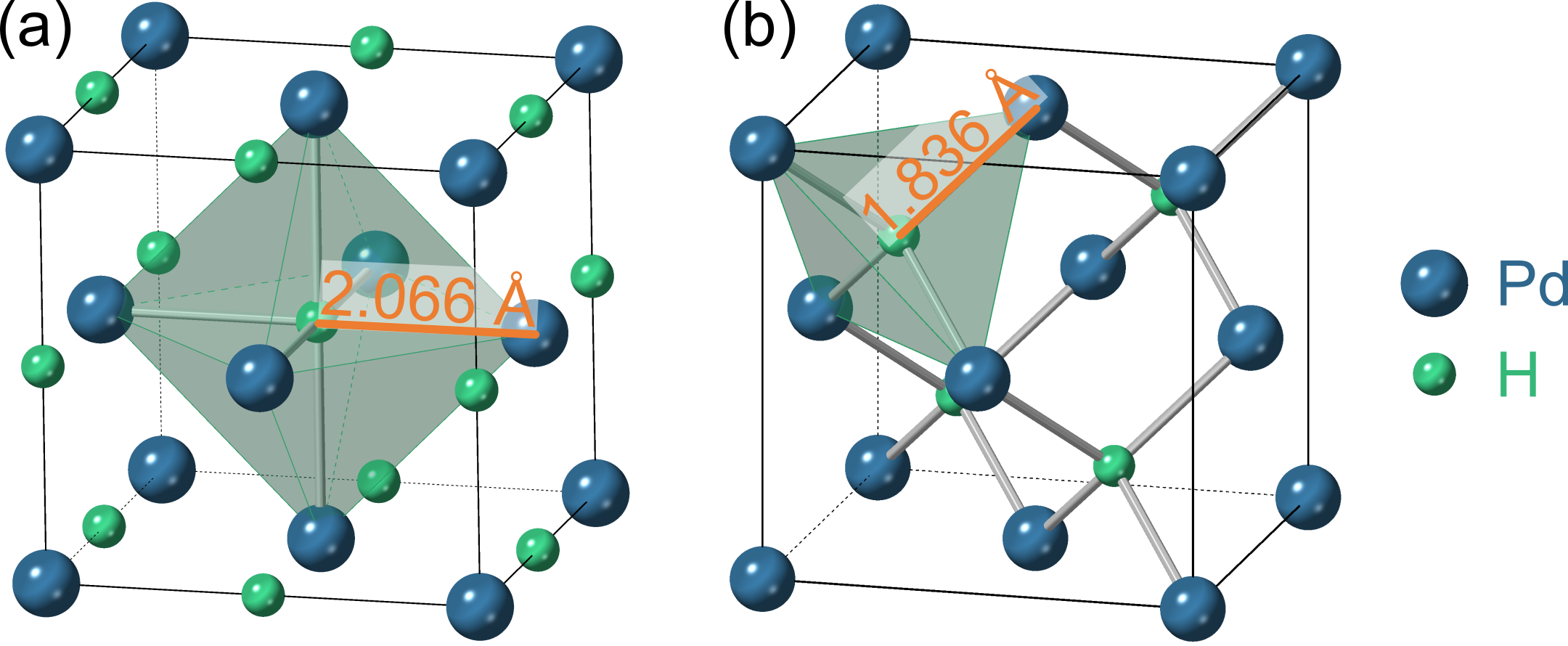}
\caption{Depiction of the unit cell for PdH$_x$ with $x$ = 1, with (a) H occupying the octahedral sites, and (b) H occupying the tetrahedral sites, including Pd--H bond lengths in \AA. Both structures are those obtained following geometry optimisations using density functional theory, where the lattice parameters were also allowed to relax. The corresponding lattice parameters are given in Table~\ref{tab:lattice}. \label{fig:structures}. For $x$ $<$ 1, the interstitial sites are only partially occupied and the lattice constant is smaller.}
\end{figure}

The palladium-hydrogen phase diagram is rather simple: at room temperature and for low hydrogen partial pressures $P_{H2}$ and, correspondingly, low hydrogen-to-metal ratios $x$, the $\alpha$-PdH$_x$ phase is formed. This phase is essentially a solid solution of hydrogen and the palladium host lattice. When $P_{H2}$ increases, $x$ increases and for $x$ $\gtrsim$ 0.02  attractive H\textendash H interactions start to dominate. This makes it possible to nucleate a high-hydrogen $\beta$-PdH$_x$ with $x$ $\gtrsim$ 0.6 that coexists for 0.02 $\lesssim$ $x$ $\lesssim$ 0.6 with the $\alpha$-PdH$_x$ phase. The $\beta$-PdH$_x$ bears the same crystal structure as the $\alpha$ phase: the Pd atoms are organised in a face-centered cubic lattice in which the hydrogen atoms occupy the octahedral interstitial sites (as depicted in Fig.~\ref{fig:structures}(a)).~\cite{flanagan1991palladium,manchester1994,lewis2000,wicke2005} The hydrogen-to-metal ratio $x$ also strongly depends on temperature according to Van 't Hoff's law: Given the negative enthalpy of formation, at a given partial hydrogen pressure, the $x$ decreases with increasing temperature.

On the theoretical side, the electronic structure of both Pd and PdH$_x$ has been the subject of research since the 1970s.~\cite{switendick1972,Faulkner1976,Zbasnik1976,Jena1979} Since then, theory has been used to offer insights into aspects including varying stoichiometries, differing structures, and octahedral \emph{vs.}\ tetrahedral H occupancies (see Fig.~\ref{fig:structures} for a depiction of both), as well as the effects of pressure, see e.g.\ Refs.~\onlinecite{Chan1983,Elsasser1991,Caputo2003,Isaeva2011,Houari2014,Yang2017,Setayandeh2021a,Setayandeh2021b,Meninno2023}.

Pd and its interaction with hydrogen have also been a fruitful playground for photoelectron spectroscopy (PES), both using UV (UPS) and X-ray (XPS) photon sources, particularly in the 1970s and 1980s, supported by the availability of high-purity single crystals and metal foils, as well as robust in-situ cleaning procedures for Pd metal.~\cite{Huefner1973,Smith1974,Demuth1977,Eberhardt1983,Antonangeli1975,Schlapbach1983,Bennett1982} More recently, a small number of studies have explored the Pd-H system using ambient pressure XPS (AP-XPS), predominantly focusing on core state spectra,~\cite{Teschner2005,Delmelle2015,Tang2022} with even fewer studies exploring Pd alloys and other metal hydrides.~\cite{White2019,Tarditi2025} A common thread of discussion throughout both the fundamental studies of the latter half of the 20th century and most recent AP-XPS work is the surface sensitivity of PES techniques and the potential influence this could have on the resulting spectra. This has left some level of uncertainty in the ability to explore bulk chemical bonding and the electronic structure of hydrides. 

The present work represents the first systematic ambient-pressure hard X-ray photoelectron spectroscopy (AP-HAXPES) study of hydrogen incorporation in Pd, utilising hard X-rays to achieve a greater probing depth and provide insight into the effect of hydrogen incorporation in the bulk. The use of hard X-rays enables the continuous application of a hydrogen pressure of 200~mbar, and in-situ heating allows for simultaneous variation of sample temperature; therefore, the changes induced by varying levels of hydrogen incorporation in Pd can be observed. Replicating comparable conditions to those achieved within the AP-HAXPES setup using structural techniques, such as X-ray diffraction and neutron reflectometry, it is possible to correlate structural changes with the chemical and electronic changes observed in AP-HAXPES. The discussion of the changes in the electronic structure induced by the incorporation of hydrogen in Pd is presented in detail, comparing relaxed lattice parameters and densities of states (DOS) from density functional theory (DFT) calculations.   

%%%%%%%%%%%%%%%%%%%%%%%%%%%%%%%%%%%%%
\section{Methods}
%%%%%%%%%%%%%%%%%%%%%%%%%%%%%%%%%%%%%

%%%%%%%%%%%%%%%%%%%%%%%%%%%%%%%%%%%%%
\subsection{Experimental}
%%%%%%%%%%%%%%%%%%%%%%%%%%%%%%%%%%%%%
\subsubsection{Sample preparation}
The samples consist of a 50~nm layer of Pd and a 4~nm Ti layer on a Si wafer. The additional 4~nm Ti adhesion layer was used to prevent delamination of the Pd from the substrate. The Si substrates (p++/B-doped 525 $\pm$ 25 $\mu$m thick Si wafers with $<$100$>$ orientation, Si-Mat, Kaufering, Germany) had a dimension of 5 $\times$ 5~mm$^2$ for the HAXPES, XRD, and XRR experiments and 3~inch (76.2~mm) in diameter and 5~mm in thickness for the neutron reflectometry experiments. The layers were produced by magnetron sputtering in an ultra-high vacuum chamber (AJA Int.) with a base pressure of $P$ $<$ 10$^{-6}$~Pa and in 0.3~Pa of Ar. The targets had a diameter of 2~inch (50.8~mm) and a purity of at least 99.99\% (Mateck, J\"{u}lich, Germany), and the distance between target and substrate was 150 mm. A schematic of the geometry of the deposition system can be found in Ref.~[\citenum{bannenberg2025palladium}]. For Pd, the deposition power was 50~W DC, yielding a deposition rate of 0.13~nm~s$^{-1}$. For Ti, the deposition power was 100~W, yielding a rate of 0.05~nm~s$^{-1}$. Before commencing the measurements, the samples were exposed to three cycles of 10~bar H$_2$ at room temperature to settle the microstructure. The substrates were rotated at 20 rotations per second to obtain a homogeneous thickness. All samples were pre-exposed to three cycles of 10~bar of hydrogen, followed by vacuum to settle the microstructure. 

To check the thickness of the Pd layer, XRR measurements were performed, which are displayed in Fig.~1 in the Supplementary Information (see below for experimental details). To fit the results, a three-layer model was used: one for the Ti adhesion layer, one for the Pd layer, and an additional layer for a surface layer (possibly PdO$_x$ or a non-Gaussian roughness), and the density of the substrate was fixed at its theoretical value of 2.6~g~cm$^{-3}$. The analysis showed a thickness of 49.8~nm for the Pd layer, a roughness of 0.7~nm, and a density of 59~atoms~nm$^{-3}$ (11~g~cm$^{-3}$). The interfacial SiO$_x$ layer was fitted to be 1.6~nm thick with a corresponding roughness of 1.2~nm. The Ti layer has a thickness of 4.3~nm, a density of 52~atoms~nm$^{-3}$ (4.1~g~cm$^{-3}$) and a roughness of 1.0~nm. As such, the densities of the layers are slightly lower than those of the bulk. 

%%%%%%%%%%%%%%%%%%%%%%%%%%%%%%%%%%%%%
\subsubsection{X-ray diffraction and reflectometry}
X-ray diffraction and reflectometry measurements, both ex-situ and in-situ, were performed using a Bruker D8 Discover (Cu K$\alpha$, $\lambda$ = 0.1542~nm) equipped with a LYNXEYE XE detector operating in 0D mode (Bruker AXS GmbH, Karlsruhe, Germany). The X-ray reflectometry measurements were performed with a G\"{o}bel mirror and a 0.1~mm exit slit on the primary and two 0.1~mm slits on the secondary side. Two measurements were performed: one for 0\textdegree\ $<$ $2\theta$ $<$ 2\textdegree\ with a 0.1~mm thick Cu attenuator and one without an attenuator for 1\textdegree\ $<$ $2\theta$ $<$ 4\textdegree. These measurements were stitched with a home-written Python code, and the data were fitted using GenX 3.6.20 to obtain estimates of the layer thickness, roughness, and density of the Pd layer. 

The X-ray diffraction experiments were also performed in a configuration with a G\"{o}bel mirror and a 1.0~mm exit slit on the primary and two 1.0~mm slits on the secondary side. Ex-situ XRD measurements were performed for 30\textdegree\ $<$ $2\theta$ $<$ 90\textdegree\ while in-situ measurements focused on 35\textdegree\ $<$ $2\theta$ $<$ 41\textdegree\ owing to the epitexture of the film in the (111) direction. The data were fitted to two pseudo-Voigt functions: one for the $\alpha$-PdH$_x$ (111) peak and one for the $\beta$-PdH$_x$. Using Bragg’s law, the out-of-plane d-spacing was calculated based on the fitted positions.

In-situ XRD and XRR measurements were performed using the same set-up as in Ref.~[\citenum{bannenberg2023completely}]. In short, this setup consists of an Anton Paar XRK900 reactor chamber (Anton Paar GmbH, Graz, Austria) connected to auxiliary pressure and temperature control equipment. Measurements as a function of temperature were performed with and without the presence of hydrogen. First, a measurement was performed at 25~\textdegree C. The sample was then heated to 200~\textdegree C under vacuum. Subsequently, the sample was exposed to $P_{H2}$ = 200~mbar, achieved by exposing the sample to $P_{tot}$ = 5,000 mbar 4.0\% H$_2$ in He ($\Delta c_{H2}$/$c_{H2}$ $<$ 2\%, Linde Gas Benelux BV, Dieren, The Netherlands). The temperature was subsequently reduced stepwise to 25~\textdegree C. At each temperature, the sample was held for 15~min to achieve (thermal) equilibrium. To compensate for the thermal expansion of the sample, measurements were also performed with the same thermal cycles under vacuum. 

%%%%%%%%%%%%%%%%%%%%%%%%%%%%%%%%%%%%%
\subsubsection{Neutron reflectometry}
In-situ neutron reflectometry experiments were performed at the time-of-flight neutron reflectometer of the TU Delft Reactor Institute~\cite{bannenberg2023ROG} using a cold neutron spectrum of 0.3 $\leq$ $\lambda$ $\leq$ 1.6 nm and a position-sensitive detector. During the measurement, the chopper was operated at a frequency of 13.2~Hz and an interdisc distance of 280~mm, resulting in a wavelength resolution of $\Delta \lambda$/$\lambda$ = 2.5\%. Measurements were performed at 4, 12, and 30~mrad incident angle. For the smallest angle, the first and second slit were set to 0.6 and 0.3~mm, respectively, and scaled with the angle for the subsequent incident angle. This resulted in a constant footprint of 65/80 $\times$ 40~mm$^2$ umbra/penumbra and an angular resolution of $\Delta \theta$/$\theta$ = 3.6\%. As such, the momentum transfer resolution was $\Delta Q$/$Q$ $\approx$ 4.5\%. The in-situ hydrogenation experiments were performed using the temperature-, flow-, and pressure-controlled cell described in Ref.~[\citenum{bannenberg2024versatile}] following the same protocol as for the in-situ XRD measurements.

The X-ray and neutron reflectometry data were fitted with GenX3,\cite{glavic2022genx}, yielding estimates for the layer thickness, density, and roughness of each layer. To compute the hydrogen-to-palladium ratio $x$, the obtained thickness $d$ and scattering length density (SLD) of the palladium layer were used:

\begin{equation}
    x = (\frac{SLD_{PdHx}}{SLD_{Pd}}\frac{d_{PdHx}}{d_{Pd}}-1)\frac{b_{Pd}}{b_H},
    \label{eq:NR}
\end{equation}

\noindent where $b_{H}$ = -3.74~fm and $b_{Pd}$ = 5.91~fm are the scattering lengths of hydrogen and palladium, respectively.~\cite{sears1992neutron} A full derivation of this formula can be found in Ref.~[\citenum{bannenberg2024structural}]. In fitting the data, a 2.5~nm thick SiO$_x$ layer was included to account for the native oxide present on the thin film, and the thickness and density of this and the Ti layer were kept constant during the analysis.

%%%%%%%%%%%%%%%%%%%%%%%%%%%%%%%%%%%%%
\subsubsection{AP-HAXPES}
%%%%%%%%%%%%%%%%%%%%%%%%%%%%%%%%%%%%%
Ambient-pressure hard X-ray photoelectron spectroscopy (AP-HAXPES) was performed on the POLARIS setup at beamline P22, PETRA III, DESY.~\cite{Amann2019,Schlueter2019} The instrument is based upon a virtual cell approach, where a local pressure cushion is formed in front of the solid sample of interest, by directing a gas stream directly onto the
flat sample surface. Under these conditions, the surrounding chamber experiences much lower pressures (on the order of a few mbar when the virtual cell is at 1~bar). Photoelectrons are collected at a distance of approx.\ 90~$\mu$m by a line array of 22~$\mu$m circular apertures matching the X-ray footprint on the sample. Experiments were conducted in total external reflection mode, with grazing incidence conditions of 0.5$^\circ$, to enhance the surface sensitity.~\cite{Amann2019} A photon energy of 4.596~keV, further referred to as 4.6~keV for simplicity, was selected using a double-bounce monochromator with Si 311 crystals. The beamline optics use a horizontally bent elliptical mirror and two vertical cylindrical focusing mirrors to achieve a beam footprint on the order of 15 $\times$ 15~$\mu$m$^2$, which was measured with a polished YAG crystal at regular intervals during the experiment. The electron analyser was used with an 800~$\mu$m curved slit and a pass energy of 200~eV for survey spectra and 100~eV for core level and valence spectra. The total energy resolution was determined from the 16/84\% width of the intrinsic Fermi edge of the Pd film sample at 200~$^\circ$C as 290~meV. 

Hydrogen gas with a purity of 5~N and a flow rate of 2.8~lpm was used for all experiments, and it was purified using a specialised gas filter (SAES Getters/Entegris). Because of the narrow gap between sample and spectrometer, an N-type thermocouple measuring the temperature was placed at the back of the sample, and the temperatures stated in the following correspond to these measurements. The temperature at the front of the sample is expected to be considerably lower due to hydrogen’s high thermal conductivity (about seven times higher than nitrogen’s). However, due to experimental limitations, it was not possible to accurately collect the temperature at the front of the sample during the experiment. Therefore, we will only refer to the minimum ($T_{min}$) and maximum ($T_{max}$) sample temperatures in the remainder of the discussion. During the AP-HAXPES measurements, the sample temperature was decreased from a nominal temperature of 200~$^\circ$C under a continuous pressure of 200~mbar \ce{H2} to a temperature of interest to obtain different hydrogen-to-metal ratios in the Pd sample. Before moving to the next temperature, the sample was first brought back to a temperature of 200~$^\circ$C.

%%%%%%%%%%%%%%%%%%%%%%%%%%%%%%%%%%%%%
\subsection{Computational}
%%%%%%%%%%%%%%%%%%%%%%%%%%%%%%%%%%%%%
Density functional theory (DFT) calculations,~\cite{Hohenberg1964,Kohn1965} including geometry optimisations and projected density of states (PDOS) calculations, were performed for Pd metal, and PdH$_x$ for $x=$0.25, 0.5, 0.75, and 1.0 for both octahedral and tetrahedral H positions. The energy for $H_2$ used for the formation enthalpy calculation was taken from previous work, wherein a 25~\AA supercell was employed.~\cite{Kalha2024} Calculations employed the plane-wave pseudopotential code CASTEP,~\cite{Clark2005} using norm-conserving pseudopotentials with 18 explicit electrons for Pd, and a kinetic energy cut-off of 1100~eV. Monkhorst-Pack $k$-point grids were used with a $12\times12\times12$ grid for Pd metal, $4\times4\times4$, $8\times8\times8$, $6\times6\times6$, and $9\times9\times9$ for octahedral PdH, PdH$_{0.75}$, PdH$_{0.5}$, and PdH$_{0.25}$, and $8\times8\times8$, $4\times4\times4$, $6\times6\times6$, and $9\times9\times9$ and for tetrahedral PdH, PdH$_{0.75}$, PdH$_{0.5}$, and PdH$_{0.25}$.~\cite{Monkhorst1976} A $12\times12\times12$ $k$-point grid sampling was subsequently employed for generating the PDOS for all systems. The PBE exchange-correlation functional was employed.~\cite{Perdew1996} Geometry optimisations were performed, imposing symmetry, with the lattice parameters being allowed to relax, where input structures for PdH$_{0.75}$, PdH$_{0.5}$, and PdH$_{0.25}$ were generated by removing H atoms from the corresponding PdH structure. The PDOS was calculated for all relaxed structures, as well as for Pd in its unrelaxed (i.e.\ experimental) structure. OptaDOS was used to post-process PDOS calculations,~\cite{Morris2014} where 0.3~eV Gaussian smearing was applied, reflecting the total experimental resolution. Calculation workflows employed the remotemanager Python package for remote job submission.~\cite{Dawson2024}

%%%%%%%%%%%%%%%%%%%%%%%%%%%%%%%%%%%%%
\subsubsection{Comparison of PDOS with VB spectra}
To directly compare the theoretically-derived PDOS with experimental VB spectra, three key steps must be taken. Firstly, the PDOS has to be broadened (see previous paragraph). Secondly, both have to be combined on a common energy axis. Here, the PDOS was aligned to the calculated Fermi energy ($E_F$) from the respective calculations, and the VB spectra were aligned to the experimentally observed Fermi cut-off. Finally, photoionisation cross-section $\sigma$ corrections have to be applied to the PDOS to scale the contributions from specific elements and orbitals. However, theoretical cross sections from e.g.\ Scofield,~\cite{Scofield1973} which are generally used for this correction are only available for states occupied in the ground state of the atom. 

For the case of Pd, the highest orbitals for which cross sections are available are the 4\textit{s}, 4\textit{p}, and 4\textit{d} states. However, the 4\textit{s} and 4\textit{p} states are shallow core levels at binding energies (BE) of approximately 87 and 51~eV, respectively, and are unlikely to contribute to states within the VB. In fact, the assumption is that any \textit{s}- and \textit{p}-type character originates from conduction states pulled below $E_F$ in metallic systems. This presents a limitation in the case of metallic Pd and Pd hydride, where contributions from the \textit{s} and \textit{p} states are predicted by theory. 

This challenge has previously been addressed for other 4\textit{d} systems, including Ag and CdO.~\cite{Panaccione2005,Mudd2014} In the present work, we take a very similar approach to these previous works and use the In~5\textit{p} and In~5\textit{s} cross sections to estimate the Pd~5\textit{s} and 5\textit{p} cross sections. Indium is chosen as it is the next element to have both of these orbitals at least partially occupied. All cross sections used as the starting point for the corrections were taken from the work of Scofield and later digitisations.~\cite{Scofield1973,Kalha2020} Tab.~1 in the Supplementary Information summarises the so-determined cross sections.

%%%%%%%%%%%%%%%%%%%%%%%%%%%%%%%%%%%%%
\section{Results and Discussion}
%%%%%%%%%%%%%%%%%%%%%%%%%%%%%%%%%%%%%

%%%%%%%%%%%%%%%%%%%%%%%%%%%%%%%%%%%%%
\subsection{Structure}\label{sec:struc}
%%%%%%%%%%%%%%%%%%%%%%%%%%%%%%%%%%%%%

To interpret the AP-HAXPES measurements, it is essential to know the phase and hydrogen-to-metal ratio in the Pd film. Unlike most typical studies on hydrides, for experimental reasons, the AP-HAXPES measurements were performed at a constant partial hydrogen pressure of $P_{H2}$ = 200~mbar and by varying the temperature. Therefore, XRD and neutron reflectometry measurements were performed using the same protocol to identify the physical state of the sample.

\begin{figure*}
    \centering
    \includegraphics[width=0.9\linewidth]{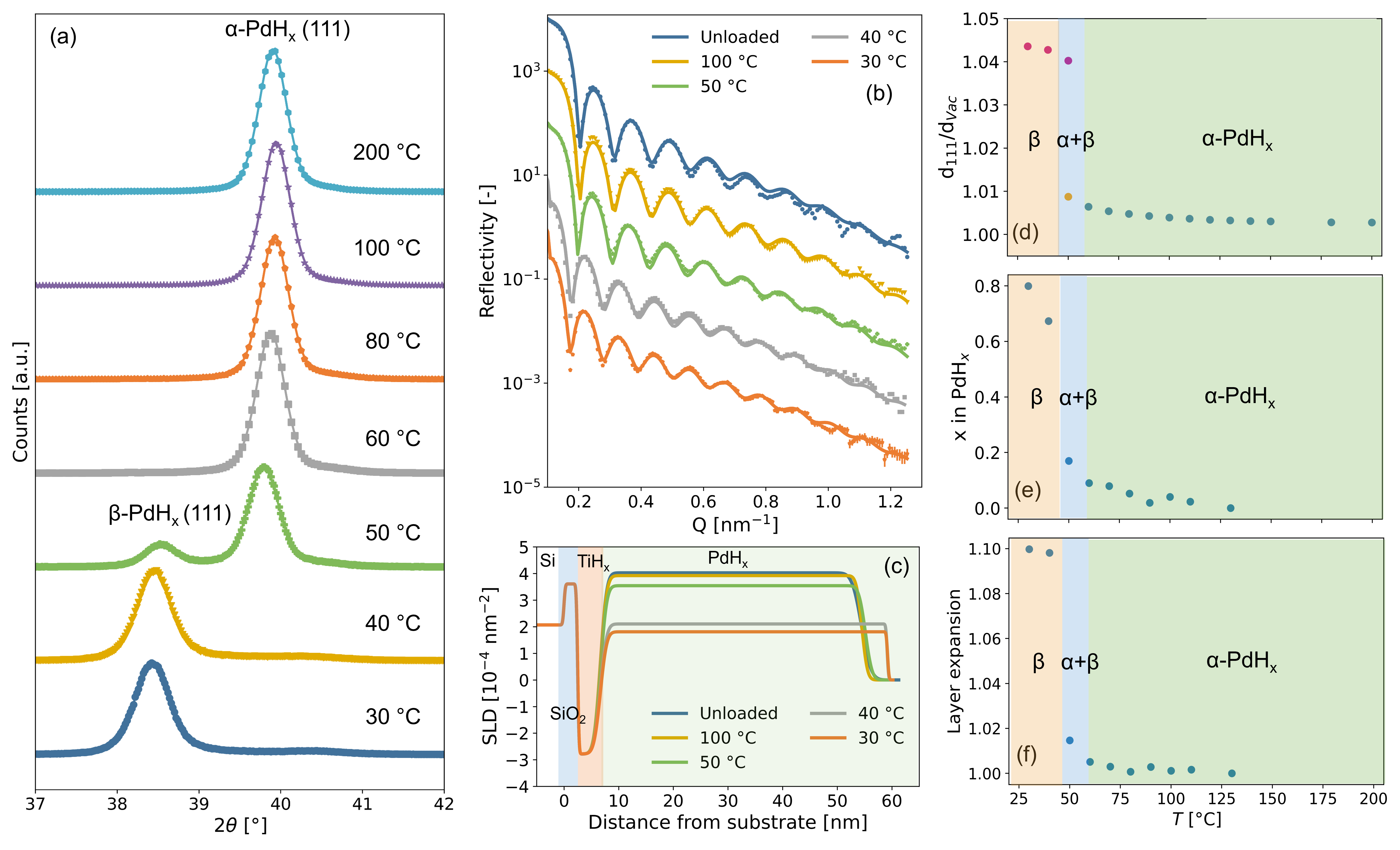}
    \caption{In-situ structural characterisation of the sample composed of a 4~nm Ti and a 50~nm Pd layer on a Si wafer. The measurements were performed by stepwise lowering the temperature from $T$ = 200~\textdegree C to the temperature indicated under $P_{H2}$ = 200~mbar. (a) Diffraction patterns around the (111) reflection (full diffraction pattern available in Fig.~2 in the Supplementary Information). The continuous lines indicate fits of a pseudo-Voigt function to the data.
    (b) Neutron reflectivity as a function of the momentum transfer $Q$. The continuous lines represent fits of a three-layer model to the data. (c) The scattering length density (SLD) profiles corresponding to the fits to the data presented in (b). Based on the SLD profiles and Equ.~1, the hydrogen-to-metal ratio and layer expansion of the PdH$_x$ layer have been determined.
    (d) Out-of-plane $d_{111}$-spacing expansion of the film due to the absorption of hydrogen. At each temperature, the $d_{111}$ spacing at $P_{H2}$ = 200~mbar is normalised to the spacing as measured in vacuum, $d_{vac}$.
    (e) Temperature dependence of the hydrogen-to-metal ratio and (f) layer expansion at $P_{H2}$ = 200~mbar.
    }
    \label{Structural}
\end{figure*}

Fig.~\ref{Structural}(a) provides the results of in-situ XRD measurements that clearly mark the $\alpha$-$\beta$ PdH$_x$ transition around $T$ = 50~\textdegree C. As the film is highly textured with $\langle 111\rangle$ in the out-of-plane direction, the following discussion focuses on this reflection. At high temperatures, the low hydrogen concentration $\alpha$-PdH$_x$ phase is formed. In this solid solution phase, hydrogen occupies the octahedral interstitial sites in the fcc palladium lattice.~\cite{worsham1957neutron} In accordance with Van 't Hoff's law, when the temperature is decreased, the diffraction peak moves towards lower angles, corresponding to the expansion of the unit cell owing to an increase in the hydrogen-to-metal ratio. At $T$ = 50~\textdegree C, a second diffraction peak emerges. This indicates the first-order phase transition to the high-concentration $\beta$-PdH$_x$ phase that has the same crystal structure as the $\alpha$-PdH$_x$ phase. When the temperature is further decreased, the fraction of the material in the $\beta$-PdH$_x$ increases at the expense of the $\alpha$-PdH$_x$, and eventually for $T \lesssim$ 40~\textdegree~C the material is entirely in its alpha phase. A further decrease in temperature results in a further increase in d-spacing until at $T$ = 30~\textdegree~C it expanded by about 4.3\% w.r.t.\ the unloaded state (from $d_{111}$ =  0.2243~$\pm$ 0.0002 to 0.2341~$\pm$ 0.0002 nm, i.e.\ $a$ = 0.3884~$\pm$ 0.0004 to 0.4055 $\pm$ 0.0004~nm assuming a cubic unit cell, which, as we will see later, is not correct). In addition, this value is also larger than the value reported in the literature of $a$ = 0.389~nm.

Figs.~\ref{Structural}(b-c) show the neutron reflectometry data, used together with the XRD diffraction patterns to construct the phase diagrams in (d-f). Herein, we present, in addition to the $d_{111}$ spacing, the hydrogen-to-metal ratio and layer expansion of the Pd thin film as a function of temperature, as determined by neutron reflectometry. In accordance with the XRD measurements and Van 't Hoff's law, reducing the temperature increases the hydrogen-to-metal ratio. Within the $\alpha$-PdH$_x$ phase, a maximum hydrogen-to-metal ratio of $x$ $\approx$ 0.1 is observed at $T$ = 60~\textdegree C, while a maximum hydrogen-to-metal ratio of $x$ $\approx$ 0.8 is observed when the material is fully in the $\beta$-phase. At this point, the layer thickness expanded by about 10\% which is a similar volumetric expansion as for bulk PdH$_x$.~ \cite{flanagan1991palladium,manchester1994,lewis2000,wicke2005} Due to the clamping of the film to the substrate, the out-of-plane lattice expansion and the resulting high in-plane stresses, the out-of-plane lattice expansion of 4.3\% is larger than for bulk palladium (around 3.7\%).~\cite{manchester1994} As the volume (layer) expansion is significantly smaller than (1.043)$^3$, it indicates that a reduced in-plane expansion and a (plastic) deformation of the palladium unit cell occurs in the thin film. Such effects have been observed before, especially if adhesion of the film to the substrate is good (see, e.g.,~\cite{laudahn1999determination,pivak2011,wagner2016,wagner2016mechanical,harumoto2017}).

Tab.~\ref{tab:lattice} shows the relaxed lattice parameters from DFT calculations for hydrogen occupation of both octahedral and tetrahedral sites. Apart from PdH$_{0.5}$, which shows a slight deviation from cubic symmetry for the octahedrally occupied structure and a larger deviation for the tetrahedrally occupied structure, all other structures remained cubic. However, under the explored conditions PdH$_{0.5}$ does not exist experimentally; instead a phase coexistence of $\alpha$-PdH$_{0.1}$ and $\beta$-PdH$_{0.6}$ is observed, and thus the theoretical PdH$_{0.5}$ is included only as a means of observing trends, rather than as a direct comparator to experiment. The lattice parameter for PdH agrees well with previous calculations, which used the PBE functional, with other functionals yielding slightly smaller lattice parameters.~\cite{Elsasser1991,Houari2014,Setayandeh2021a,Setayandeh2021b} Furthermore, as expected, the lattice parameter monotonically increases with increasing H for both octahedrally- and tetrahedrally-occupied structures. Specifically, there is an 11.7\% (17.9\%) increase in cell volume going from Pd to PdH$_{0.75}$ for the octahedrally (tetrahedrally) occupied structure. The expansion of the unit cell for PdH$_{0.75}$ matches the experimental value of $\approx$10\% reasonably well for the case of the octrahedral sites being partially occupied. At the same time, a much larger discrepancy is found for the tetrahedral case. As such, this is consistent with the idea that hydrogen occupies the octahedral sites in thin film Pd, just like for bulk Pd.

\begin{table}[ht]
\centering
%\begin{threeparttable}
\caption{Relaxed lattice parameters and relative cell volume increases from DFT, for which all angles were constrained to be 90$^\circ$.}
\label{tab:lattice}
\begin{tabular*} {\linewidth}{l @{\extracolsep{\fill}} rrrr}
\hline \hline
%multirow{2}{*}{
& $a$ (\AA) & $b=c$ (\AA) & $V$(\AA$^3$) & $\frac{V-V_{\mathrm{Pd}}}{V_{\mathrm{Pd}}}$  (\%)\\
\cline{1-1}\cline{2-2}\cline{3-3}\cline{4-4}\cline{5-5}\\[-2.5ex]
Pd & 3.946 & 3.946 & 61.422 & -\\
\cline{1-1}\cline{2-2}\cline{3-3}\cline{4-4}\cline{5-5}\\[-2.5ex]
\textbf{Octahedral}\\
PdH$_{0.25}$ & 4.003 & 4.003 & 64.120 & 4.392 \\
PdH$_{0.5}$ & 4.046 & 4.054 & 66.493 & 8.255\\
PdH$_{0.75}$ & 4.094 & 4.094 & 68.599 & 11.684\\
PdH & 4.132 & 4.132 & 70.554 & 14.866\\
\cline{1-1}\cline{2-2}\cline{3-3}\cline{4-4}\cline{5-5}\\[-2.5ex]
\textbf{Tetrahedral} \\
PdH$_{0.25}$ & 4.027	& 4.027 &	65.286	& 6.290\\
PdH$_{0.5}$ & 4.048	& 4.204	& 68.886	& 12.150\\
PdH$_{0.75}$ & 4.168	& 4.168 & 72.419	& 17.903\\
PdH & 4.239	& 4.239	& 76.193 & 24.047\\
 \hline \hline
\end{tabular*}
%\end{threeparttable}
\end{table}

%%%%%%%%%%%%%%%%%%%%%%%%%%%%%%%%%%%%%
\subsection{Chemical Bonding and Electronic Structure}
%%%%%%%%%%%%%%%%%%%%%%%%%%%%%%%%%%%%%

\begin{figure*}
    \centering
    \includegraphics[width=0.8\linewidth]{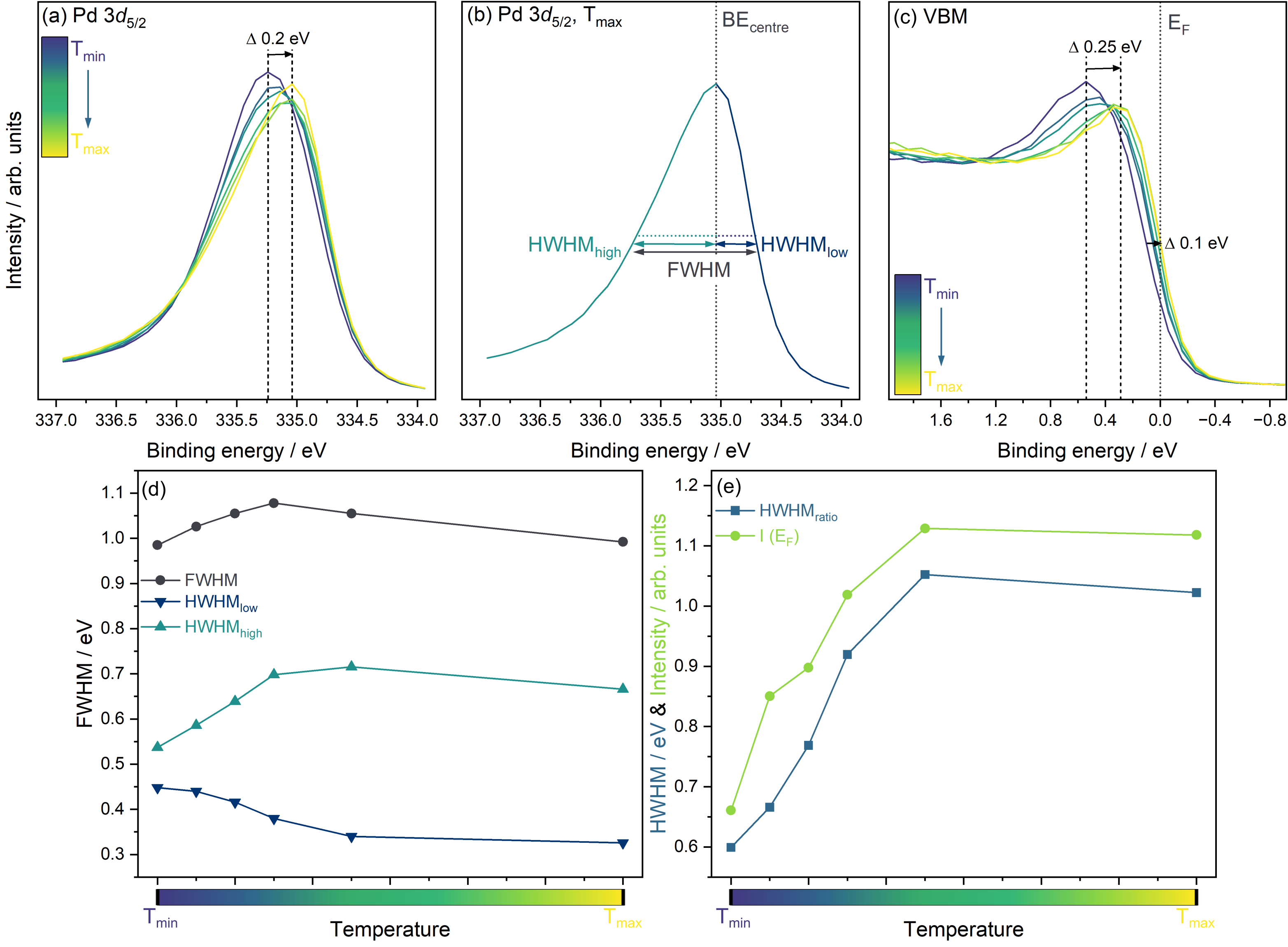}
    \caption{AP-HAXPES data collected during the heating of a hydrided Pd film under 200~mbar \ce{H2}. (a) Pd~3\textit{d}$_{5/2}$ core state spectra. Dashed lines and the arrow illustrate shifts in the main peak position. (b) Pd~3\textit{d}$_{5/2}$ AP-HAXPES core state spectrum after heating to the maximum temperature (T$_{max}$) with depiction of the defined peak widths, where FWHM is the total full width at half maximum, and HWHM\textsubscript{low} and HWHM\textsubscript{high} are the half width at half maximum for the low and high binding energy (BE) side of the binding energy centre (BE\textsubscript{centre}). (c) Valence band maximum (VBM) spectra. Dashed lines and arrows illustrate shifts in the VBM position. The position of the Fermi energy $E_F$ at 0~eV is indicated. (d) Extracted Pd~3\textit{d}$_{5/2}$ peak widths. (e) Ratio of the two Pd~3\textit{d}$_{5/2}$ HWHMs compared to the single point signal intensity at $E_F$, I ($E_F$).}
    \label{XPS1}
\end{figure*}

AP-HAXPES was used to investigate the chemical bonding and electronic structure of the Pd film upon hydrogen incorporation. Survey spectra were collected at each temperature point, showing all expected Pd core lines (see Fig.~3 in the Supplementary Information). In addition, a minute amount of carbon was also observed (see Fig.~4 in the Supplementary Information). Both the Pd core and valence state PES are sensitive to the incorporation of hydrogen. After the preparation of the initial hydrided Pd film, measurements were performed at several temperatures by cooling down the film from 200 $^\circ$C under partial hydrogen pressure to a temperature of interest, were a lower temperature results in more hydrogen absorbed by the film (Fig.~\ref{Structural}). This resulted in clear differences between the spectra. The Pd~3\textit{d}$_{5/2}$ core line, shown in Fig.~\ref{XPS1}(a), shifts continuously to lower binding energy (BE), from 335.3 to 335.0~eV, with the maximum difference between the minimum (T$_{min}$) and maximum (T$_{max}$) sample temperatures being 0.20~eV. This is in good agreement with previous PES work, which observed shifts between 0.15 and 0.18~eV (with an error of between 0.05 and 0.1~eV given).~\cite{Schlapbach1983,Bennett1982,Teschner2005}

A change in the spectral lineshape accompanies this shift in peak position. Both the full width at half maximum (FWHM) and peak asymmetry change gradually with changes in the amount of hydrogen present in the Pd film. In general, the asymmetry in the core-state spectra of metallic systems arises from the interaction of the positive core hole created in the photoemission process with the mobile conduction electrons.~\cite{Huefner1975} As the conduction electron population decreases, as is the case upon the formation of a hydride from pure Pd, the coupling between the core hole and the conduction electrons weakens, leading to a decrease in asymmetry. Doniach and \v{S}unji\'c developed a formal model describing this behaviour.~\cite{Doniach1970} Fitting the Pd line shape is notoriously difficult even with the model developed by Doniach and \v{S}unji\'c due to the long-range decay of the higher binding energy (BE) tail.~\cite{Huefner1975} Coupled with the need to measure restricted BE ranges in AP-HAXPES experiments to allow economic use of beamtime, instead of choosing a peak fitting approach to quantify these changes, a simpler peak asymmetry analysis was undertaken. Fig.~\ref{XPS1}(b) shows the Pd~3\textit{d}$_{5/2}$ spectrum for the sample collected at $T_{max}$ and the approach chosen for the analysis. For each spectrum, the total FWHM as well as the peak centre (BE\textsubscript{centre}) was extracted. The FWHM was then subdivided into a lower BE half width at half maximum (HWHM\textsubscript{low}) representing the symmetric half of the spectrum and a higher BE half width at half maximum (HWHM\textsubscript{high}) representing the asymmetric part. 

As noted, the core spectra asymmetry is directly connected to conduction electrons, and the valence states in the AP-HAXPES experiments exhibit a clear Fermi cut-off at 0~eV, as expected for metallic systems. Fig.~\ref{XPS1}(c) shows an expanded view of the valence band maximum (VBM), where feature I shifts from 540~meV above $E_F$ at $T_{min}$ to 260~meV at $T_{max}$. In addition, the position of the VBM shifts to lower BE by 100~meV over the same temperature range.

Fig.~\ref{XPS1}(d) shows the three peak widths extracted from the series of Pd~3\textit{d}$_{5/2}$ spectra shown in Fig.~\ref{XPS1}(a). Focusing on the total FWHM, this increases by 0.1~eV with a decrease in hydrogen content, in agreement with the literature.~\cite{Bennett1982} Interestingly, the behaviour of HWHM\textsubscript{low} and HWHM\textsubscript{high} diverges. The lower BE onset of the core level spectra becomes steeper with an increase in temperature and the resulting loss of hydrogen, leading to a drop in HWHM\textsubscript{low} by 0.12~eV. In contrast to this, the HWHM\textsubscript{high} representing the asymmetry, increases with a loss of hydrogen by 0.18~eV, resulting in an initial net increase in FWHM. The loss of asymmetry in PdH compared to Pd has been well documented in the literature.~\cite{Antonangeli1977,Bennett1982,Teschner2005} 

To explore the correlation between the core spectral shape and the population at $E_F$, Fig.~\ref{XPS1}(e) compares the DOS at $E_F$, represented by the single point signal intensity at $E_F$, I ($E_F$) (see Fig.~\ref{XPS1}(c)) and the ratio of the two HWHM, sometimes referred to as the asymmetry index. The trends across the experimental temperature range in both are comparable, further emphasising the link between the conduction state population and core line shape, in explicit agreement with the understanding of the influence of hydrogen incorporation on the electronic structure of Pd.

Beyond the changes discussed at the VBM and $E_F$, several differences are observed across the valence region. To analyse these in more detail, PDOS were calculated for Pd and PdH$_x$. Unweighted PDOS for both the octahedrally- and tetrahedrally-occupied structures are shown in Figs.~5 and 6 in the Supplementary Information, respectively. In line with the literature,~\cite{Houari2014,Setayandeh2021b} the octahedrally- and tetrahedrally-occupied structures show qualitative disagreements. Notably, the octahedrally-occupied structure gives rise to a much better agreement with experiment, as will be discussed in the following, in line with the better agreement with respect to experimental cell volume for the octahedrally-occupied structure. As such, only the octahedrally-occupied PDOS will be considered in detail. In addition, the PDOS for Pd was calculated using both the DFT-relaxed structure and the experimental structure, as shown in Fig.~7 in the Supplementary Information. While there is an increase in overall bandwidth for the experimental (unrelaxed) structure, the differences are relatively small, with no significant differences in shape. 

One potential limitation of the theoretical approach is that a cubic unit cell was assumed, which conflicts with the experimental observations discussed in Section~\ref{sec:struc}. However, the impact on the lattice parameters is small, and as evidenced by the comparison with the experimental Pd lattice parameter, it is unlikely to lead to significant differences in the PDOS. Other factors which can influence the accuracy of the PDOS include the choice of exchange-correlation functional; however, previous work has found that the differences between functionals, as well as pseudopotentials, are small.~\cite{Setayandeh2021b}

\begin{figure*}
    \centering
    \includegraphics[width=0.55\linewidth]{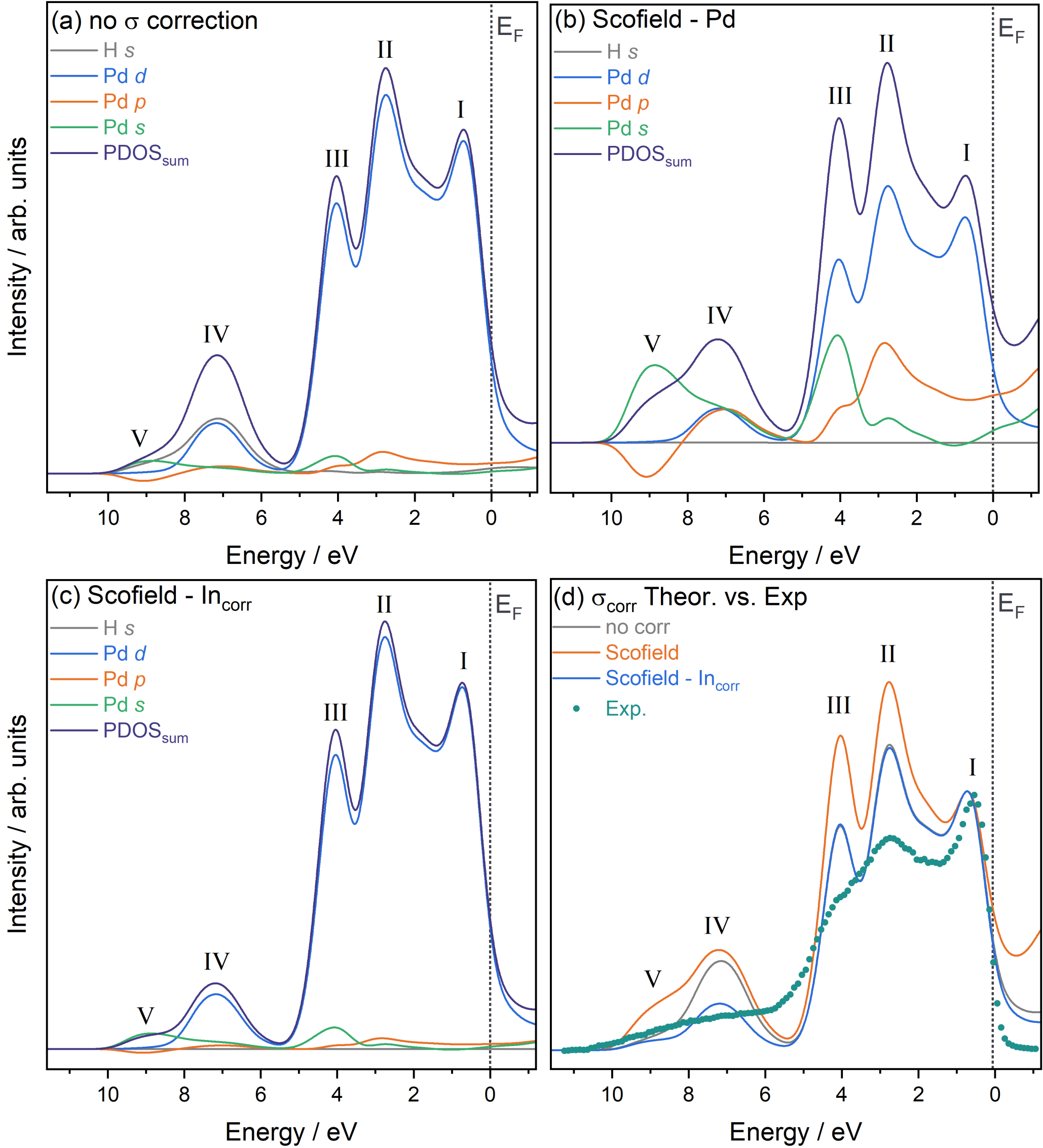}
    \caption{Comparison of PDOS for PdH\textsubscript{0.75} with different photoionisation cross-section, $\sigma$, correction strategies, including (a) PDOS without correction, (b) using Scofield $\sigma$ for Pd, (c) using Scofield $\sigma$ with In correction, and (d) comparison of all three theoretical PDOS correction approaches with the experiment at highest hydrogen loading. The position of the Fermi energy $E_F$ at 0~eV is indicated in all Subfigures, and Roman numerals are used to indicate the main spectral features observed. In (d), PDOS and VB spectrum are normalised to the height of feature I.}
    \label{XPS_CS}
\end{figure*}

As noted in the Methods section, due consideration must be given to the approach used to correct theoretical PDOS with photoionisation cross-sections. Fig.~\ref{XPS_CS}(a) shows the uncorrected PDOS for PdH$_{0.75}$ as an example. As expected, this is dominated by Pd~\textit{d} states with only minor contributions from Pd~\textit{s} and \textit{p} states as well as H~\textit{s}. After applying Pd cross-sections from Scofield (see Fig.~\ref{XPS_CS}(b)), the Pd~\textit{s} and \textit{p} contributions are exaggerated, leading to a nonsensical representation of the valence electronic structures. By applying the indium correction to the Scofield cross-sections (see Fig.~\ref{XPS_CS}(c)), a much better scaling is achieved, resulting in good comparability to the experimental VB spectra (see Fig.~\ref{XPS_CS}(d)). Therefore, the In-corrected cross-sections are chosen for all further comparisons in this work.\par

\begin{figure*}
    \centering
    \includegraphics[width=0.8\linewidth]{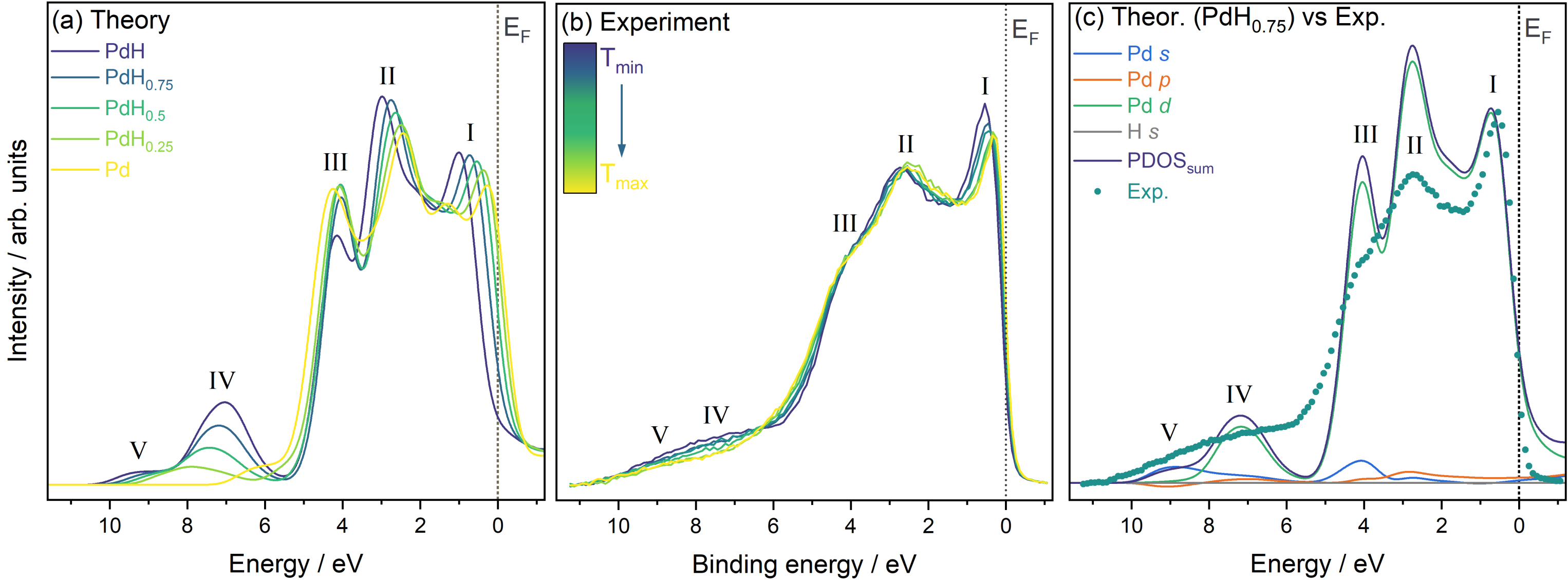}
    \caption{Comparison of AP-HAXPES valence band (VB) spectra with calculated PDOS after one-electron photoionisation cross-section correction, including (a) photoionisation cross-section corrected sums of theoretical PDOS for the Pd to PdH series, (b) AP-HAXPES VB spectra collected after exposure to 200~mbar \ce{H2} followed by heating to 200~$^\circ$C, and (c) comparison of the calculated PDOS for PdH\textsubscript{0.75} using the Scofield In $\sigma$ correction and the AP-HAXPES VB spectrum with the highest hydrogen loading. The position of the Fermi energy $E_F$ at 0~eV is indicated in all Subfigures, and Roman numerals are used to indicate the main spectral features observed. In (c), PDOS and VB spectrum are normalised to the height of feature I.}
    \label{XPS2}
\end{figure*}

The cross-section corrected PDOS and the experimental VB spectra of Pd with varying levels of hydrogen incorporation both exhibit five main features, indicated with Roman numerals I - V in Figs.~\ref{XPS2}(a) and (b). The energy axis from theory and experiment are aligned so that their respective $E_F$ positions fall at 0~eV. No further adjustment was necessary. Whilst there remains some discrepancy in the relative intensities of the features between theory and experiment, the relative energy positions and total width agree very well (in contrast to the tetrahedrally-occupied structure, see Fig.~8 in the Supplementary Information). The total width of the experimental VB is 4.8$\pm$0.1~eV. As hydrogen is introduced into Pd, the width is known to decrease, with P.~A.~Bennett and J.~C.~Fuggle previously reporting a reduction of the total width of the \textit{d} band by 10\% between Pd and PdH\textsubscript{0.8}. The experimental difference observed in the present work is 0.38~eV. This narrowing arises in part from the Pd lattice expansion that occurs upon hydrogen incorporation, leading to a reduction in Pd--Pd overlap, which is clearly visible in the changes in the calculated electron densities of the PdH$_x$ system (see Figs.~9 and 10 in the Supplementary Information). Consequently, the Pd~\textit{d} states narrow, which dominate the valence band spectra.

The individual features, particularly features I and II, also exhibit discernible BE shifts in both theory and experiment upon hydrogen incorporation. The position of feature I increases by 0.24~eV upon hydrogen incorporation, with its position changing from 0.54~eV at $T_{min}$ to 0.30~eV $T_{max}$. This matches the predicted theoretical positions for PdH$_{0.5}$ (0.54~eV) and Pd (0.28~eV). Feature II shows a comparable total experimental BE position change of 0.28~eV with its position changing from 2.72~eV at $T_{min}$ to 2.44~eV $T_{max}$. This again agrees well with the predicted position for PdH$_{0.75}$/PdH$_{0.5}$ (2.77 and 2.64~eV, respectively) and Pd (2.43~eV). Feature III does not exhibit obvious trends, with the exception of the Pd being at clearly higher BE compared to PdH$_x$. Finally, the theoretically predicted position of Feature IV changes by 0.8~eV from 7.1~eV in PdH to 7.9~eV in PdH$_{0.25}$, but any differences in the experimental position of these features cannot be resolved due to the increased lifetime broadening.

Fig.~\ref{XPS2}(c) shows the direct comparison of the PDOS for PdH$_{0.75}$ and the VB at the lowest experimental temperature $T_{min}$, corresponding to the highest hydrogen loading. As expected, the VB (features I - IV) is dominated by Pd~$d$ states, whilst some intermixing with \textit{p} and \textit{s} states occurs toward the middle (feature II) and bottom (feature III) of the VB, respectively. Below the main VB, features IV and V, located between 6 and 10~eV, are hydrogen-induced states. These states are absent in pure Pd and only appear upon the incorporation of H, an observation clearly confirmed by both theory and experiment (see Figs.~\ref{XPS2}(a) and (b), respectively). This is also in good agreement with previous observations from UPS,~\cite{Eastman1971, Demuth1977} and XPS.~\cite{Antonangeli1975,Bennett1982,Schlapbach1983} The appearance of these lower BE states (IV and V) can be understood as follows. When hydrogen starts to occupy sites within the Pd structure, its 1\textit{s} orbital overlaps with Pd states, leading to a stabilisation of these Pd--H bonding states compared to the purely metallic Pd states, leading to a lowering of their BE.

The enthalpy of formation $\Delta H_f$ is a key characteristic to evaluate the ability of a metal to incorporate hydrogen. As to a first approximation the loss in entropy of hydrogen formation is 130~J~K$^{-1}$ mol$_{H2}^{-1}$, a larger negative enthalpy of formation indicates that the hydride can be formed at lower hydrogen partial pressures. The enthalpy can be obtained directly from theoretical calculations, using the expression $\Delta H_f = 2 \times (E_{PdH} - (E_{Pd} + \frac{E_{H_2}}{2}))$, where $E_{PdH}$ is the energy per formula unit of the hydride, $E_{Pd}$ is the energy per atom of bulk Pd metal, $E_{H_2}$ is the energy per molecule of H$_2$, and the factor of 2 accounts for the conversion from kJ mol$_{H}^{-1}$ to kJ mol$_{H_2}^{-1}$. The resulting $\Delta H_f$ for octahedral and tetrahedral occupation in bulk \ce{PdH_{0.75}} are -34.16 and -51.32~kJ mol$_{H_2}^{-1}$. This agrees well with previous values reported in the literature for PBE,~\cite{Yang2017,Setayandeh2021b} but deviates somewhat from observed experimental values,~\cite{Griessen2016} while previous work has also shown a strong sensitivity to the choice of exchange correlation functional.~\cite{Houari2014,Setayandeh2021b} In addition, earlier measurements on comparable thin film samples to those used in the present work gave a value of -42$\pm$2~kJ~mol\textsuperscript{-1} based on optical measurements.~\cite{Bannenberg2025} The discrepancy between the formation enthalpies and expected stabilities of the tetrahedrally- and octahedrally-occupied structures is due to the lack of including the phonon energy: the order was found to be reversed for a range of exchange-correlation functionals once the zero-point energies were included.~\cite{Setayandeh2021b}

Griessen and Driessen proposed an alternative approach to predicting $\Delta H_f$ for a specific metal host. They defined an empirical linear relationship between $\Delta H_f$ and a characteristic energy, $\Delta E$, in the electronic structure of the metal:~\cite{Griessen1984} 

%%%%%%%%%%%%%%%%%%%%%%%%%%%%%%%%%%%%%%%%%%
\begin{equation}\label{Driessden}
    \Delta H_f = \frac{n_s}{2}(\alpha \Delta E + \beta),
\end{equation}
%%%%%%%%%%%%%%%%%%%%%%%%%%%%%%%%%%%%%%%%%%
%
\noindent where $\Delta E=E_s-E_F$, $E_F$~=~0~eV, $E_s$ is the centre of the lowest conduction band of the metal and equivalent to the energy at which the integrated density of states ($\int$DOS) of the metal is equal to 0.5$n_s$, $n_s$ is the number of electrons per atom in the lowest \textit{s}-like conduction band of the metal, $\alpha=29.62$ kJ/eV~mol~H, and $\beta=-135$~kJ/mol~H. We have previously shown that this approach can be improved by using the position of the lowest energy \textit{s}-dominated feature of the valence band of the hydride instead of the integrated DOS of the metal in the cases of \ce{TiH_x} and \ce{YH_x}.~\cite{Kalha2024} Due to the overlap with the strong Pd~4$d$ states, this feature cannot be directly observed in HAXPES for PdH; however, it can be extracted from the calculated PDOS. By inspecting the PDOS two low-binding energy features with $s$-character can be identified for \ce{PdH_{0.75}} with octahedral H occupation at 2.73 and 4.08~eV, with the latter having a higher DOS. Based on these energies, $\Delta H_f$ of -108.4 and -28.5~kJ mol$_{H_2}^{-1}$ result. The higher energy and density feature at 4.08~eV, therefore, provides better agreement with the DFT-predicted and previously reported experimental values. As in our previous work on \ce{TiH_x} and \ce{YH_x}, this approach provides less negative values when compared to DFT. In comparison, the value for tetrahedral H occupation in bulk \ce{PdH_{0.75}} determined by the same approach is -19.2~kJ mol$_{H_2}^{-1}$ based on a single $s$-character feature at 4.23~eV, consistent with the DFT results.

\section{Conclusions}

This work establishes direct links between hydrogen incorporation, structural expansion, chemical bonding, electronic structure modifications, and the hydride enthalpy of formation in Pd thin films by combining AP-HAXPES with in-situ XRD, neutron reflectometry, and DFT. The results demonstrate that the structural changes occurring upon hydrogen incorporation reduce Pd-Pd orbital overlap and induce hybridisation of Pd and H states, consequently leading to changes in Pd core state energies and a narrowing of the valence band, while simultaneously suppressing metallic screening at the Fermi level, influencing the asymmetry of the core state spectra. The excellent agreement between experiment and DFT models for octahedral occupation confirms the dominant hydrogen occupancy in bulk-like PdH. Crucially, the emergence of hydrogen-induced bonding states below the valence band provides direct spectroscopic evidence of Pd--H interactions in the bulk, thereby addressing long-standing questions regarding the potential limitations of surface-sensitive methods. The identification of $s$-state contributions to the valence states allows for the deduction of values for the enthalpy of formation of the hydride. More broadly, this study highlights AP-HAXPES as a powerful probe of bulk chemical bonding in metal hydrides under realistic conditions, opening new opportunities for resolving the electronic structure of complex hydrogen–metal systems beyond PdH.

\section*{Acknowledgements}

We are grateful for computational support from the UK national high-performance computing service, ARCHER2, for which access was obtained via the UKCP consortium and funded by EPSRC grant ref EP/X035891/1. We acknowledge DESY (Hamburg, Germany), a member of the Helmholtz Association HGF, for the provision of experimental facilities. Parts of this research were carried out at beamline P22. Beamtime was allocated for proposal H-20010087.

\bibliography{refs} 
\bibliographystyle{apsrev4-1}

\end{document}

\title{Influence of Hydrogen-Incorporation on the Bulk Electronic Structure and Chemical Bonding in Palladium\\
Supplementary Information}

\author{L.~J.~Bannenberg}
 \affiliation{Faculty of Applied Sciences, Delft University of Technology, Mekelweg 15, 2629 JB Delft, The Netherlands.}

\author{F.~Garcia-Martinez}
\affiliation{Photon Science, Deutsches Elektronen-Synchrotron DESY, 22607 Hamburg, Germany.}

\author{P.~L\"{o}mker}
\affiliation{Department of Physics, Stockholm University, 10691 Stockholm, Sweden.}
\affiliation{Wallenberg Initiative Materials Science for Sustainability, Department of Physics, Stockholm University, 114 28 Stockholm, Sweden.}
\affiliation{Photon Science, Deutsches Elektronen-Synchrotron DESY, 22607 Hamburg, Germany.}

\author{R.~Y.~Engel}
 \affiliation{Department of Physics, Stockholm University, 10691 Stockholm, Sweden.}
 \affiliation{Wallenberg Initiative Materials Science for Sustainability, Department of Physics, Stockholm University, 114 28 Stockholm, Sweden.}

\author{C.~Schlueter}
\affiliation{Photon Science, Deutsches Elektronen-Synchrotron DESY, 22607 Hamburg, Germany.}

\author{H.~Schreuders}
\author{A.~Navarathna}
 \affiliation{Faculty of Applied Sciences, Delft University of Technology, Mekelweg 15, 2629 JB Delft, The Netherlands.}

\author{L.~E.~Ratcliff}%
\affiliation{Centre for Computational Chemistry, School of Chemistry, University of Bristol, Bristol BS8 1TS, United Kingdom.%\\
}%
\affiliation{Hylleraas Centre for Quantum Molecular Sciences, Department of Chemistry, UiT The Arctic University of Norway, N-9037 Troms\o{}, Norway.}
 
\author{A.~Regoutz}
  \email{anna.regoutz@chem.ox.ac.uk}
\affiliation{Department of Chemistry, University of Oxford, Inorganic Chemistry Laboratory, Oxford OX1 3QR, United Kingdom.}
 \affiliation{Department of Chemistry, University College London, London WC1H 0AJ, United Kingdom.}

\maketitle

\cleardoublepage

%%%%%
\section{Structural Characterisation}
%%%%%

\begin{figure}[ht!]
    \centering
    \includegraphics[width=1\linewidth]{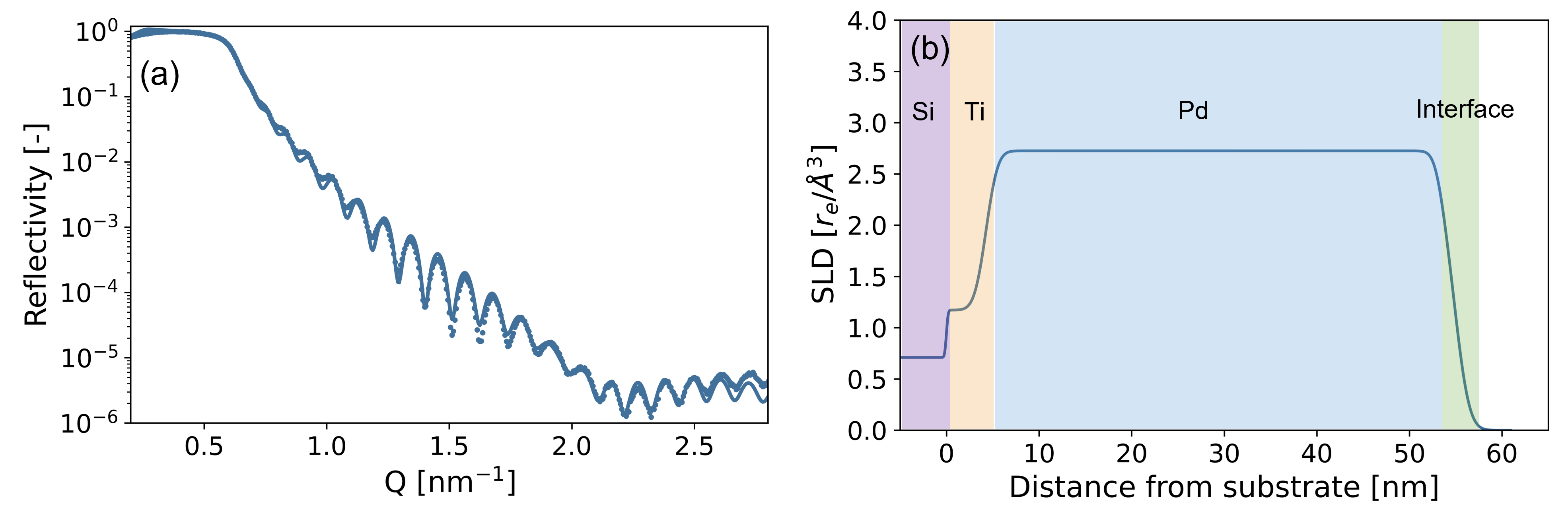}
    \caption{X-ray reflectometry (XRR) results of the sample composed of a Ti and Pd layer. (a) X-ray reflectometry data as a function of the momentum transfer $Q$. The points indicate the measured data, while the continuous line represents a fit to the data. (b) Corresponding scattering length density profile (SLD) as a function of the distance from the substrate and obtained from fitting the experimental data of (a) to a model. The SLD is, in the case of X-rays, roughly proportional to the mass density of the film.}
    \label{XRR}
\end{figure}

\begin{figure}[ht!]
    \centering
    \includegraphics[width=0.55\linewidth]{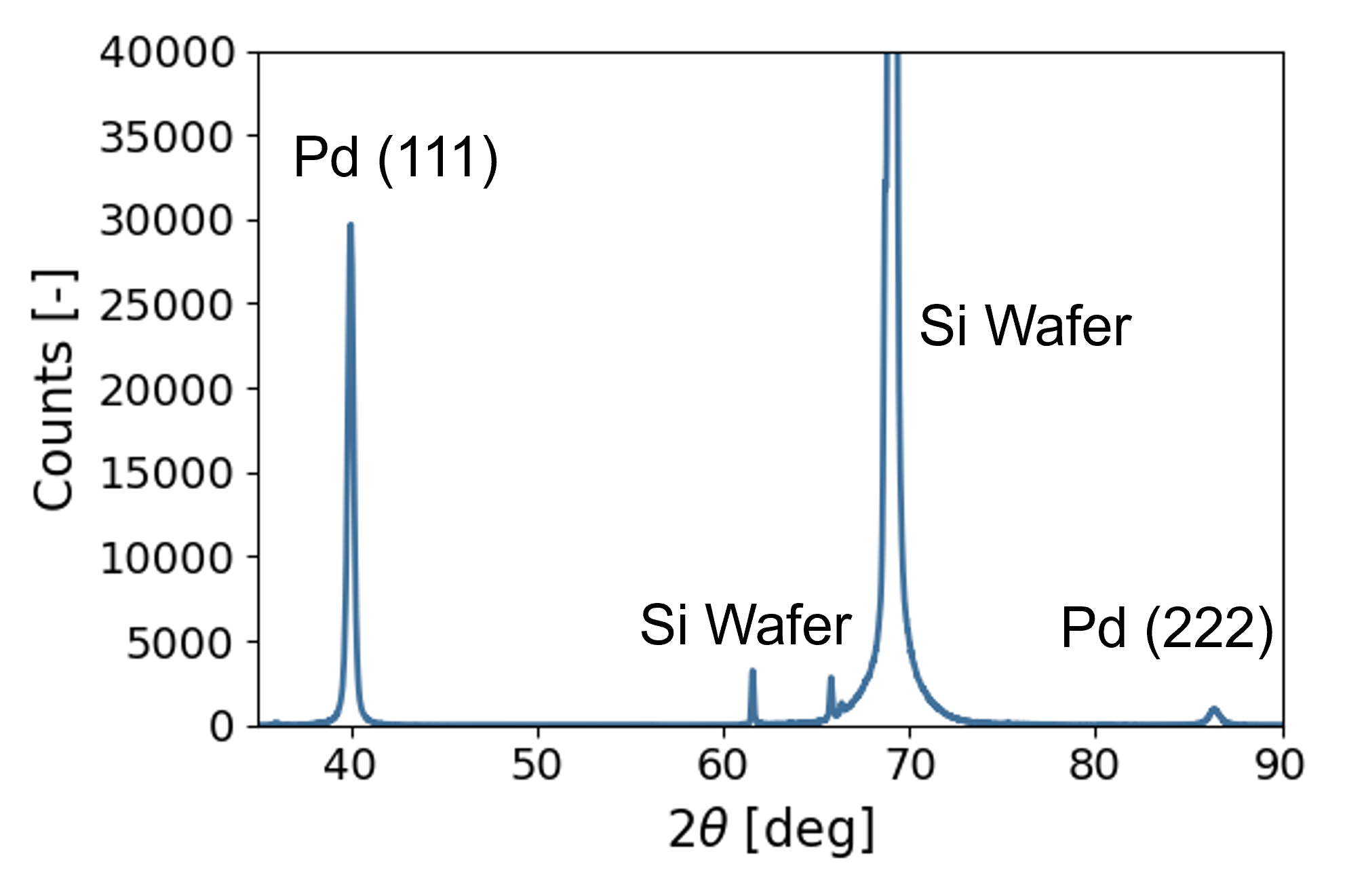}
    \caption{Ex-situ X-ray diffraction measurements of the sample composed of a 4~nm Ti and a 50~nm Pd layer on a Si wafer.}
    \label{XRDexsitu}
\end{figure}

\cleardoublepage

%%%%%
\section{AP-HAXPES \& Theory}
%%%%%

\begin{table}[ht]
\centering
\caption{One-electron corrected photoionisation cross sections, $\sigma$, and number of electrons, n(e\textsuperscript-), of Pd~\textit{s}, Pd~\textit{p}, Pd~\textit{d}, and H~\textit{s} valence states. Values from Scofield from Ref. XX are interpolated from 4 and 5~keV listed values for the experimental photon energy of 4.596~keV, with the lowest Pd~\textit{s} and \textit{p} orbitals being the deeper, occupied 4\textit{s} and 4\textit{p} states. The “Scofield In correction” approach is similar to that implemented in a previous paper on W metal, XX where the unoccupied 5\textit{s} and 5\textit{p} cross sections are estimated extrapolating from the respective orbital cross sections for In.}
\label{tab:crossec}
\begin{tabular}{lrrrrrr}
\hline \hline
\textbf{Scofield}& Pd~4\textit{s} & Pd~4\textit{p}\textsubscript{1/2} & Pd~4\textit{p}\textsubscript{3/2} & Pd~4\textit{d}\textsubscript{3/2} & Pd~4\textit{d}\textsubscript{5/2}  & H~\textit{s}\\
\hline
$\sigma$ (4~keV) & 647	& 526 & 479 & 81 & 75 & 0 \\
$\sigma$ (5~keV) & 415 & 304 & 273 & 38 & 35 & 0 \\
$\sigma$ (4.596~keV) &	553.35 & 436.16 & 395.62 & 63.75 & 58.77 & 0.04\\
n(e\textsuperscript{-}) & 2 &	2 &	4 & 4 &	6 &	2 \\
\hline
\textbf{Scofield In correction}& In~5\textit{s} & In~5\textit{p}\textsubscript{1/2} & In~5\textit{p}\textsubscript{3/2} & Pd~4\textit{d}\textsubscript{3/2} & Pd~4\textit{d}\textsubscript{5/2}  & H~\textit{s}\\
\hline
$\sigma$ (4~keV) & 81	& 37 &	30 &	81 &	75 &	0 \\
$\sigma$ (5~keV) & 53 & 22 &	18 &	38 &	35 &	0 \\
$\sigma$ (4.596~keV) &	69.29 & 30.84 &	25.26 &	63.75 &	58.77 &	0.04 \\
n(e\textsuperscript{-}) & 2 &	0.33 &	0.67 & 4 &	6 &	2 \\
 \hline \hline
\end{tabular}
\end{table}

\begin{figure}[ht!]
    \centering
    \includegraphics[width=0.55\linewidth]{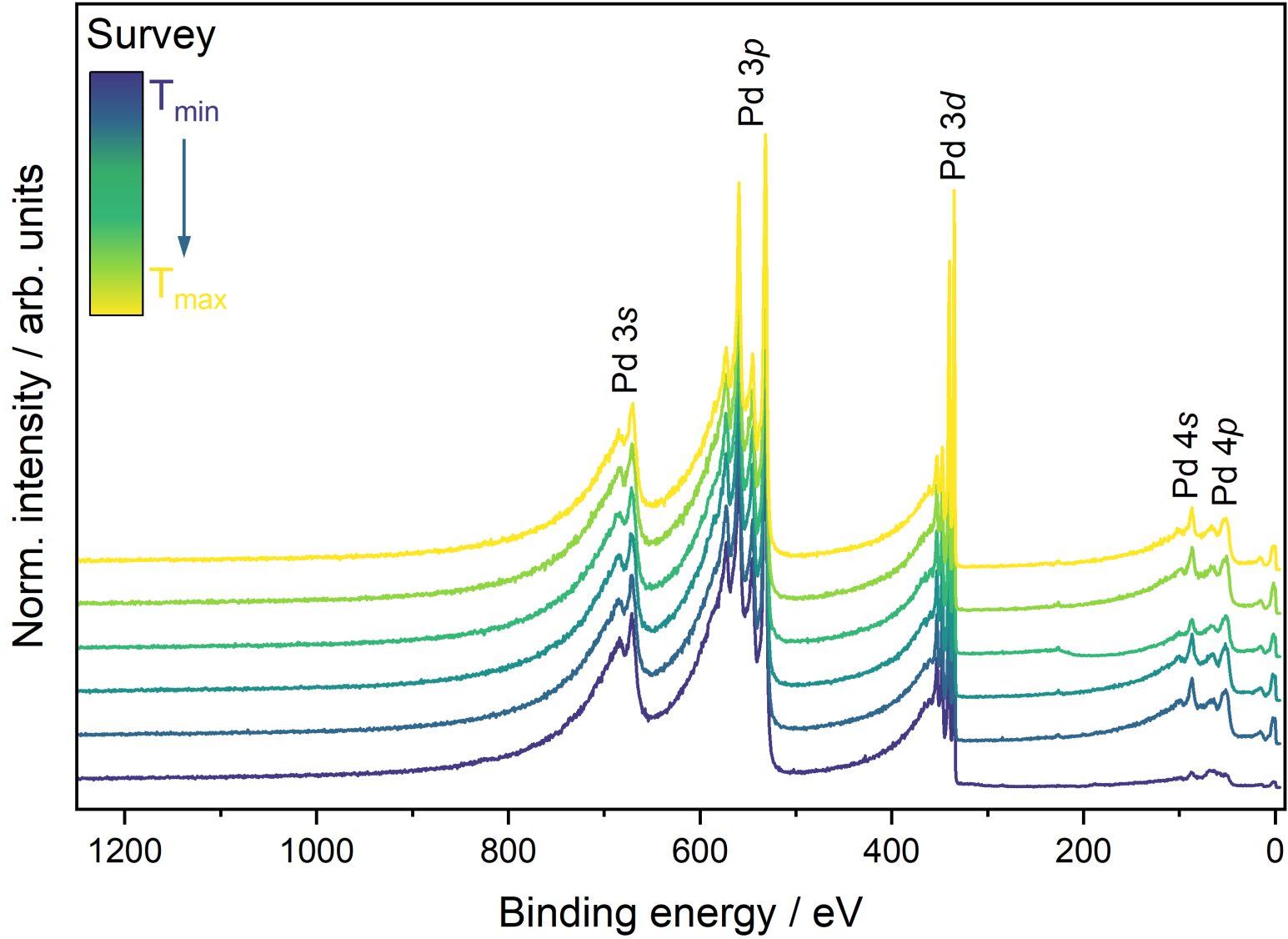}
    \caption{AP-HAXPES survey spectra collected after exposure to 200~mbar \ce{H2}, cooling to room temperature, followed by heating. All major core levels are indicated. The spectra are normalised (0,1) and stacked.}
    \label{survey}
\end{figure}

\begin{figure}[ht!]
    \centering
    \includegraphics[width=0.35\linewidth]{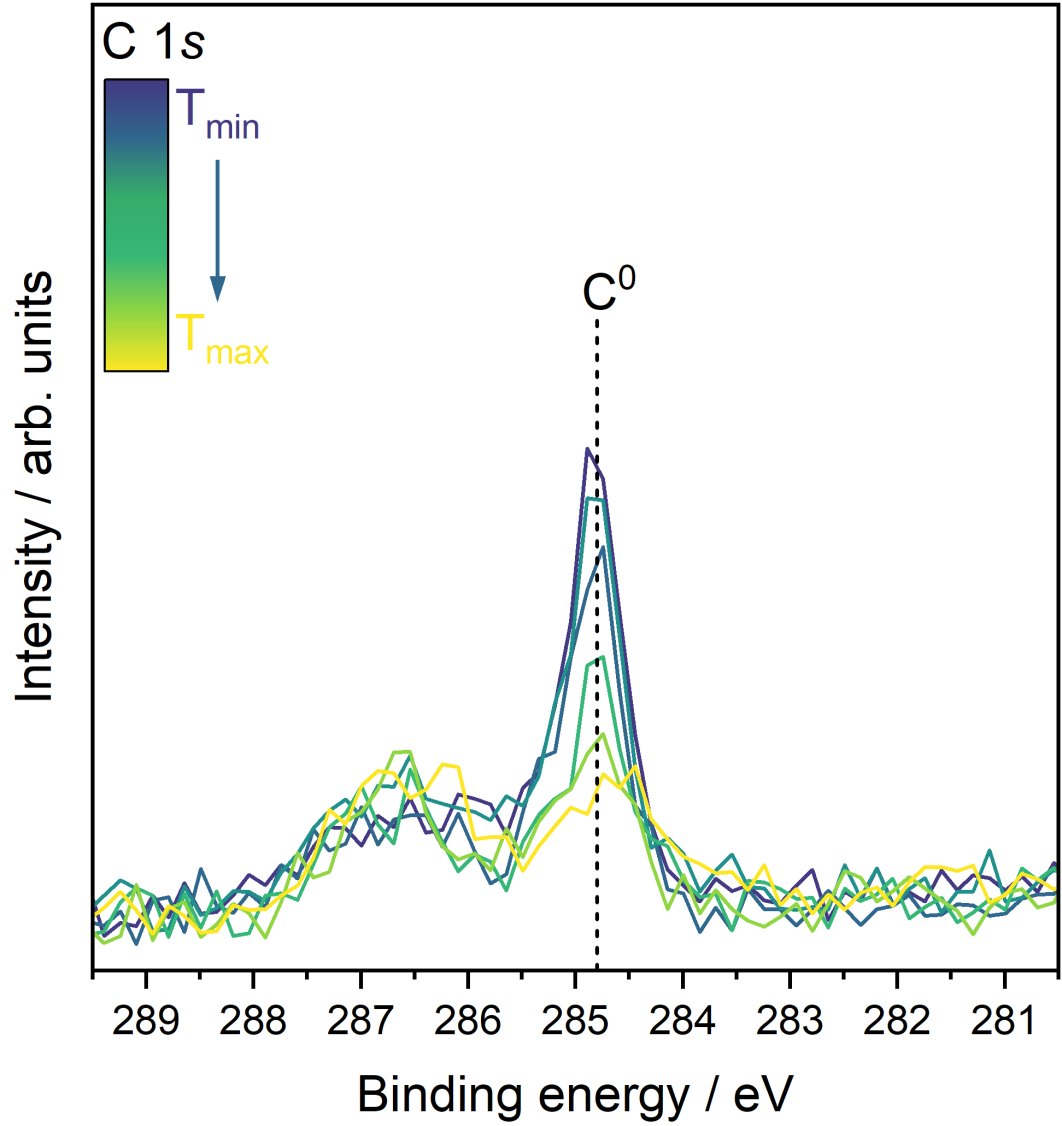}
    \caption{AP-HAXPES C~1\textit{s} spectra collected after exposure to 200~mbar \ce{H2}, cooling to room temperature, followed by heating.}
    \label{C1s}
\end{figure}

\begin{figure}[ht!]
    \centering
    \includegraphics[width=0.8\textwidth]{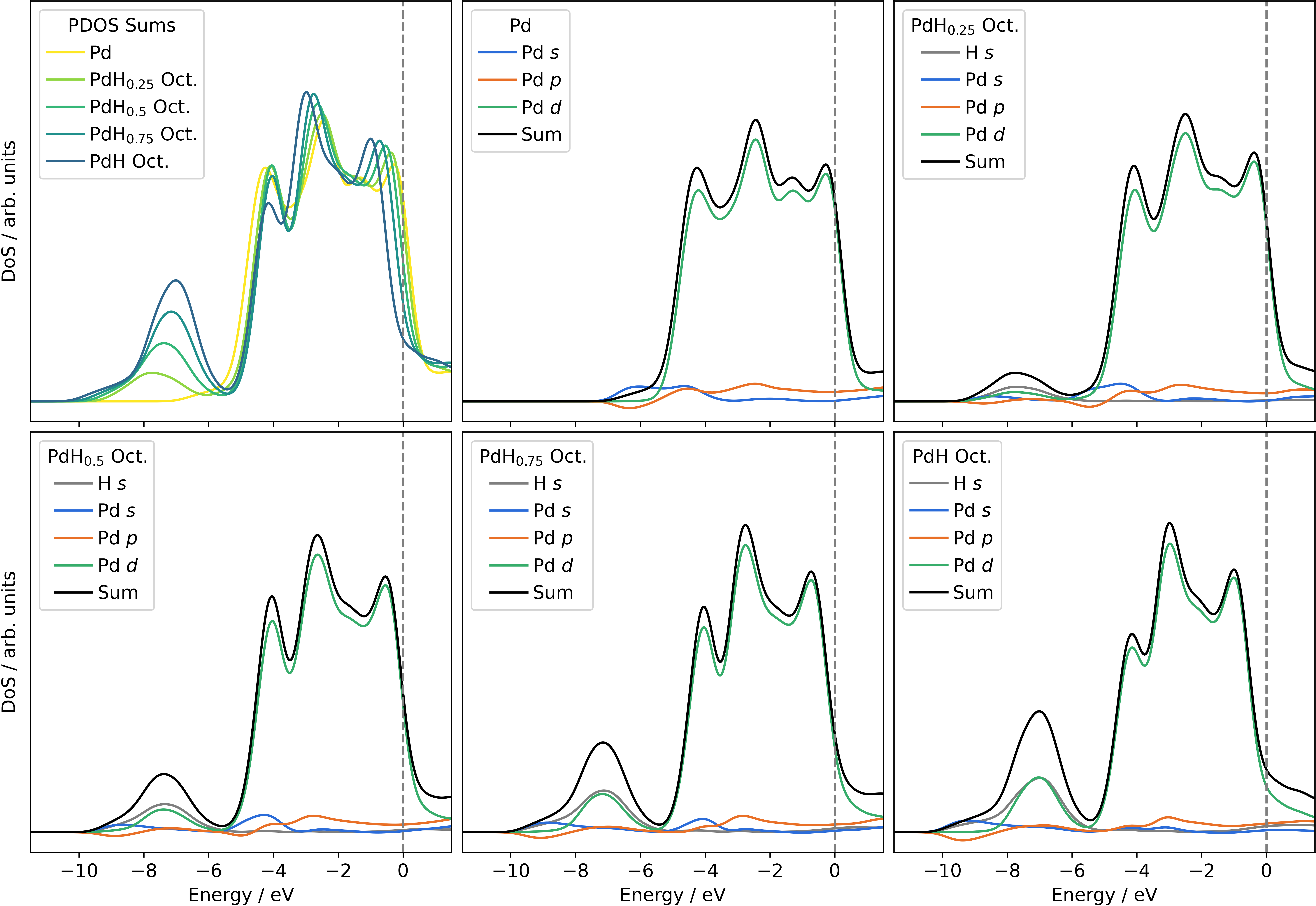}
    \caption{Projected density of states (PDOS) for the different H concentrations, with H occupying the octahedral sites. The upper left plot shows a comparison of the sum of the PDOS for all systems, while the remaining plots show the PDOS for individual systems. The PDOS have been aligned to the calculated Fermi energy, which is indicated with a dashed vertical line.}
    \label{unweighted_pdos}
\end{figure}

\begin{figure}[ht!]
    \centering
    \includegraphics[width=0.8\textwidth]{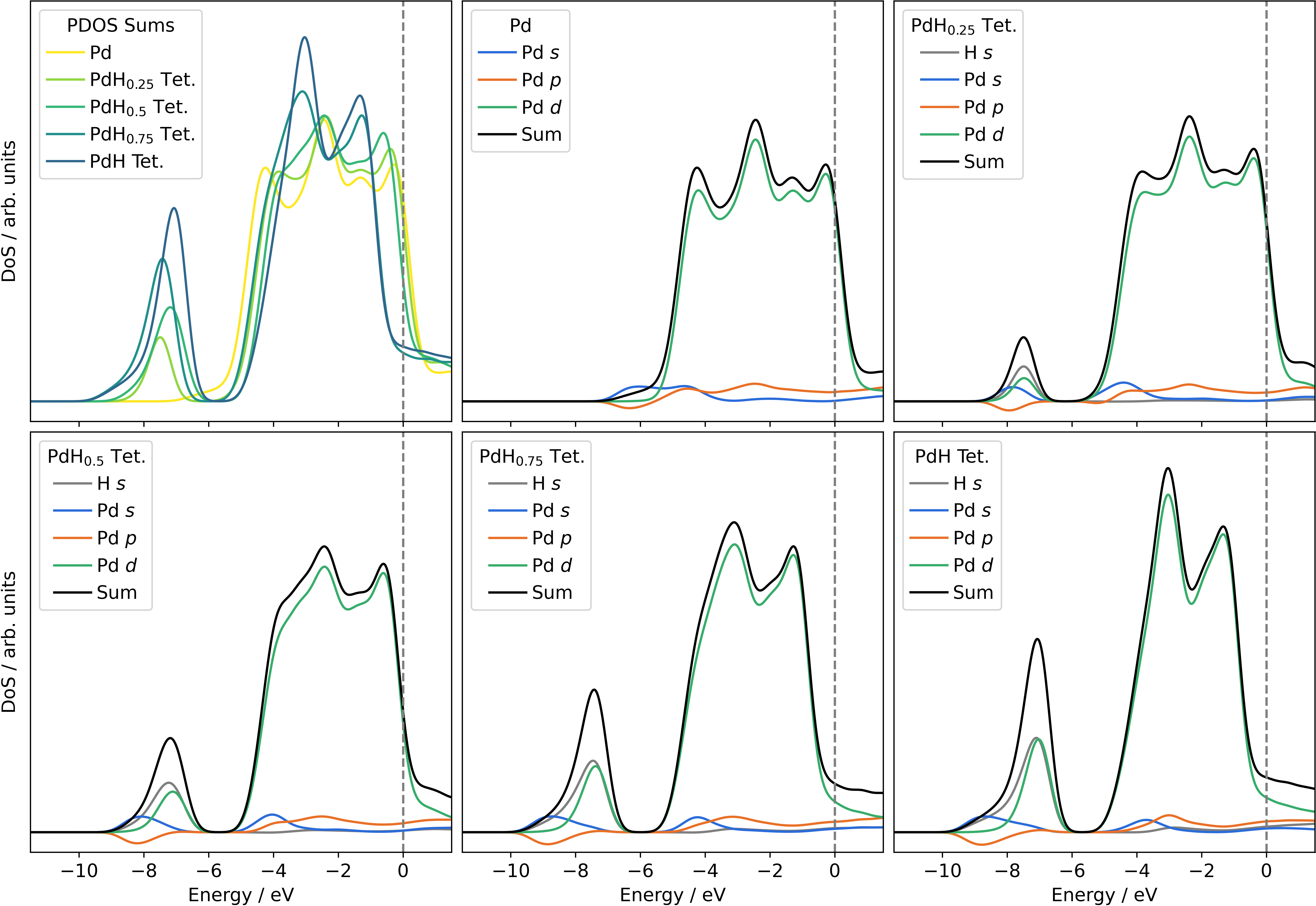}
    \caption{Projected density of states (PDOS) for the different H concentrations, with H occupying the tetrahedral sites. The upper left plot shows a comparison of the sum of the PDOS for all systems, while the remaining plots show the PDOS for individual systems. The PDOS have been aligned to the calculated Fermi energy, which is indicated with a dashed vertical line.}
    \label{unweighted_pdos_tet}
\end{figure}

\begin{figure}[ht!]
    \centering
    \includegraphics[width=0.8\textwidth]{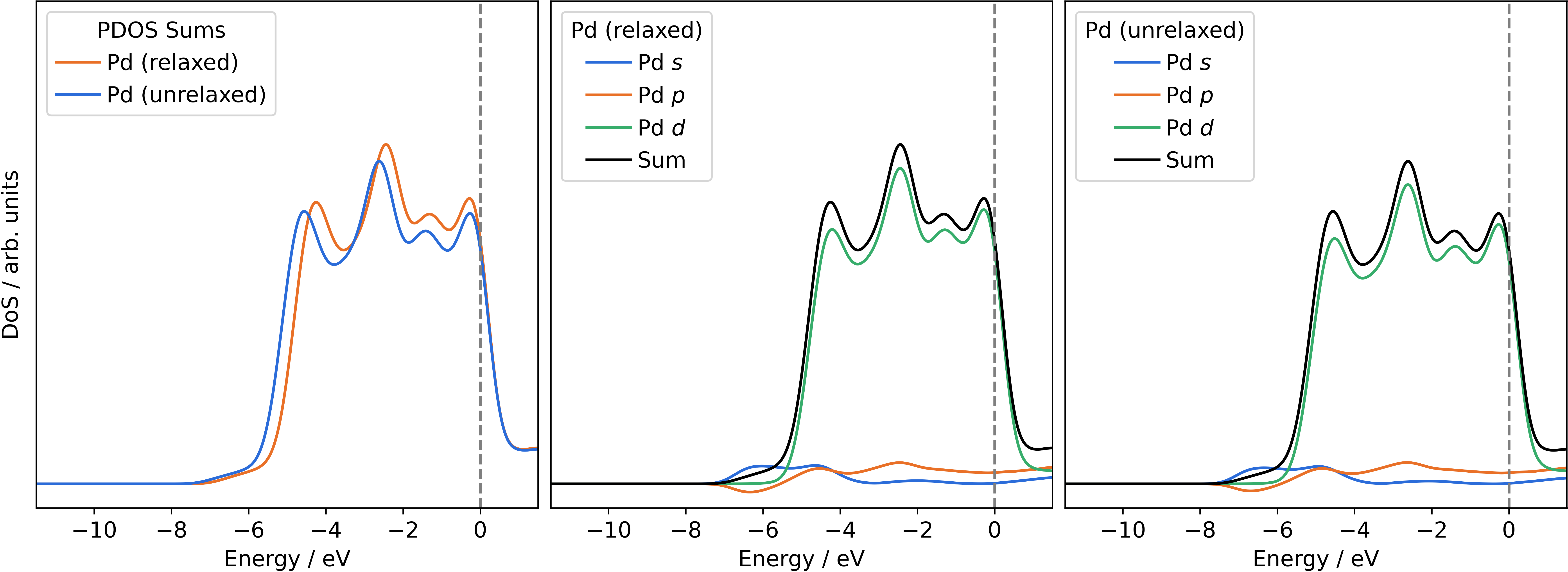}
    \caption{Projected density of states (PDOS) for Pd with both its experimental (`unrelaxed') and DFT-relaxed lattice parameter. The left plots show a comparison of the sum of the PDOS for both systems, while the remaining plots show the PDOS for individual systems.  The PDOS have been aligned to the calculated Fermi energy, which is indicated with a dashed vertical line.}
    \label{pdos_expansion}
\end{figure}

\clearpage

\begin{figure}[ht!]
    \centering
    \includegraphics[width=0.8\textwidth]{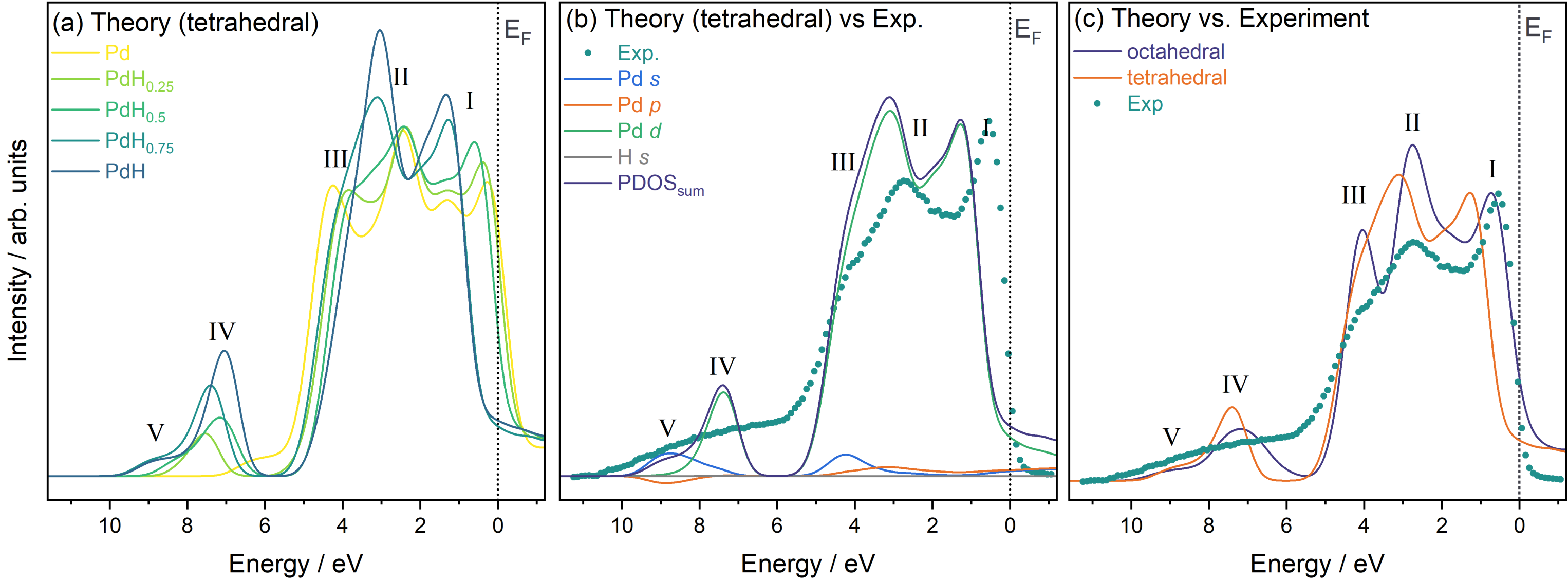}
    \caption{Comparison of PDOS for hydrogen occupying octahedral or tetrahedral sites, including (a) photoionisation cross-section corrected sums of theoretical PDOS for the Pd to PdH series assuming tetrahedral occupation, (b) comparison of the calculated PDOS for PdH\textsubscript{0.75} and the AP-HAXPES VB spectrum with the highest hydrogen loading, and (c) a comparison of the same experimental dataset with PDOS assuming either octahedral or tetrahedral occupation. All PDOS are corrected using the Scofield In $\sigma$ correction. The position of the Fermi energy $E_F$ at 0~eV is shown in all subfigures and Roman numerals are used to indicate the main spectral features observed. In (c) PDOS and VB spectrum are normalised to the height of feature I.}
    \label{tet}
\end{figure}

\begin{figure*}[ht!]
    \centering
 \begin{subfigure}[t]{0.32\linewidth}
    \centering
    \includegraphics[width=1.0\linewidth]{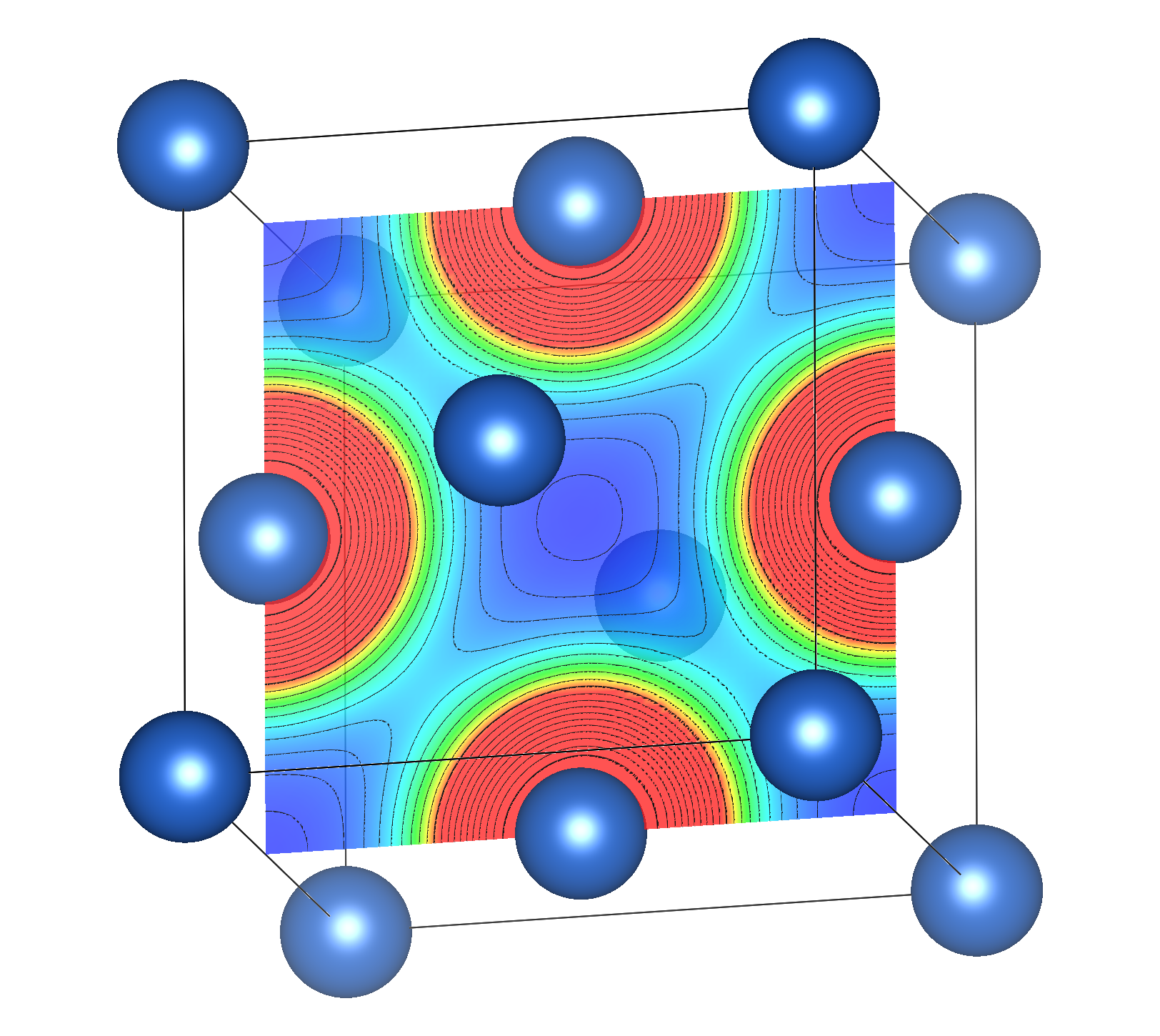}
    \caption{Pd}
\end{subfigure}
 \begin{subfigure}[t]{0.32\linewidth}
    \centering
    \includegraphics[width=1.0\linewidth]{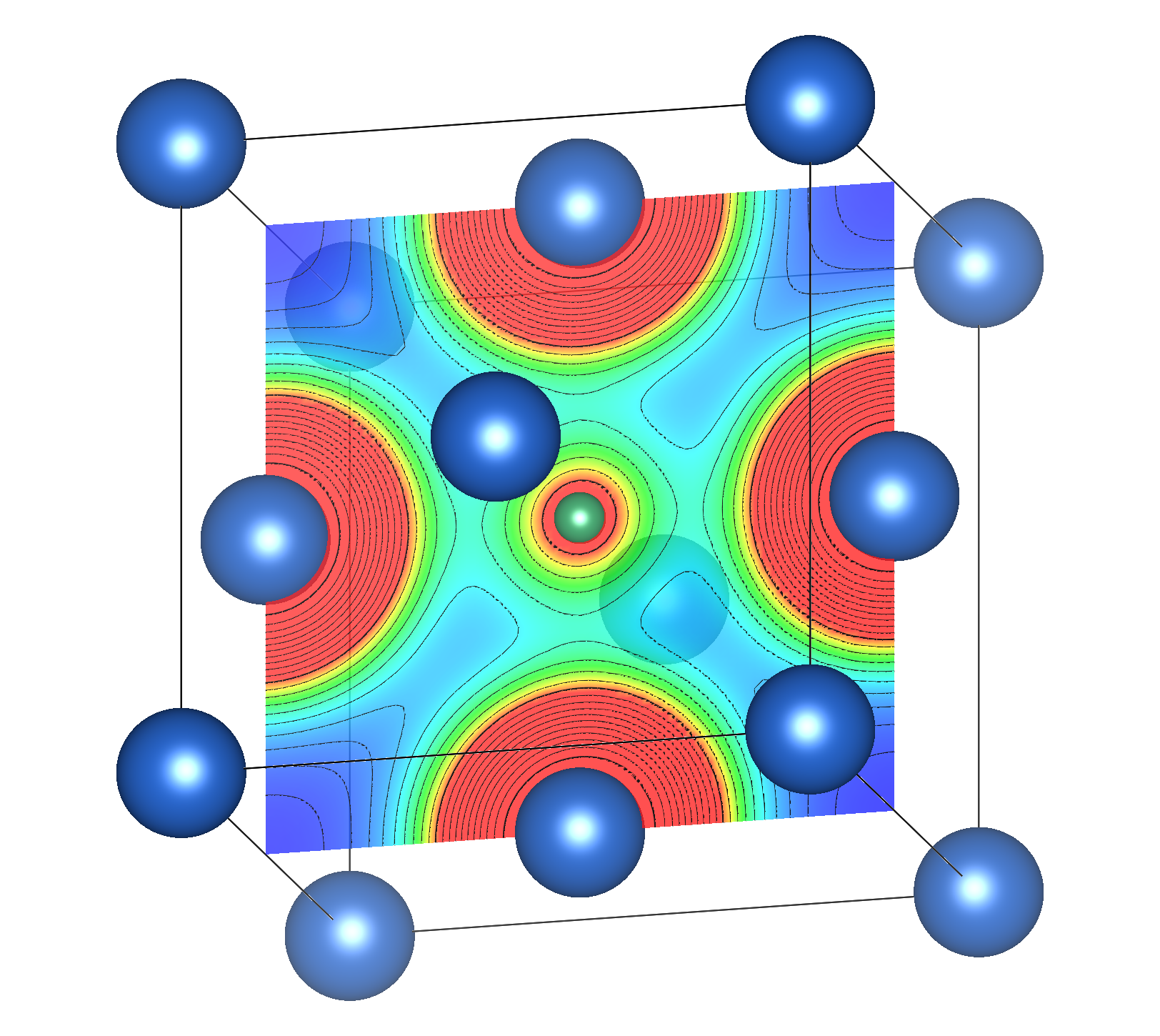}
    \caption{PdH$_{0.25}$}
\end{subfigure}
 \begin{subfigure}[t]{0.32\linewidth}
    \centering
    \includegraphics[width=1.0\linewidth]{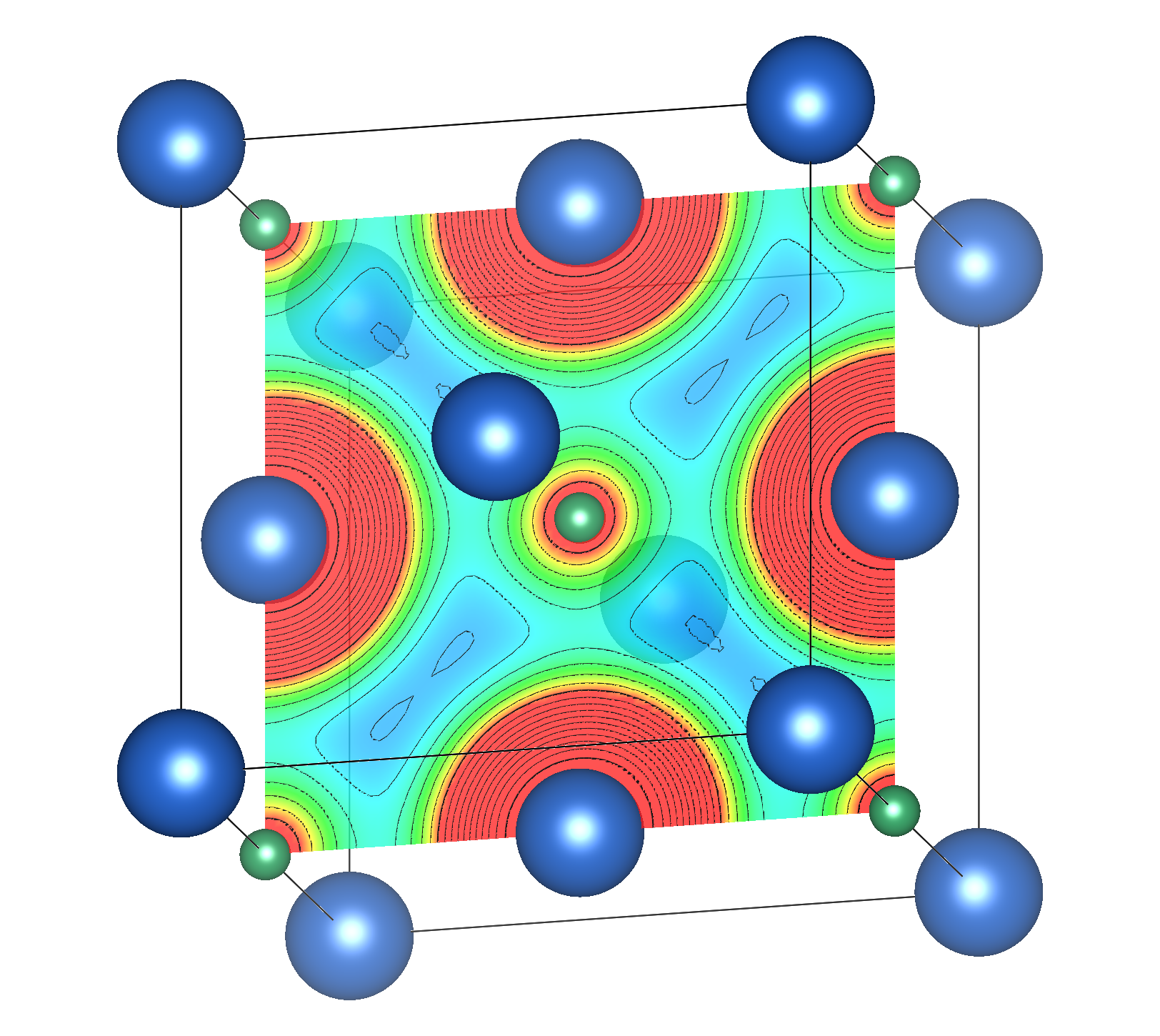}
    \caption{PdH$_{0.5}$}
\end{subfigure}
 \begin{subfigure}[t]{0.32\linewidth}
    \centering
    \includegraphics[width=1.0\linewidth]{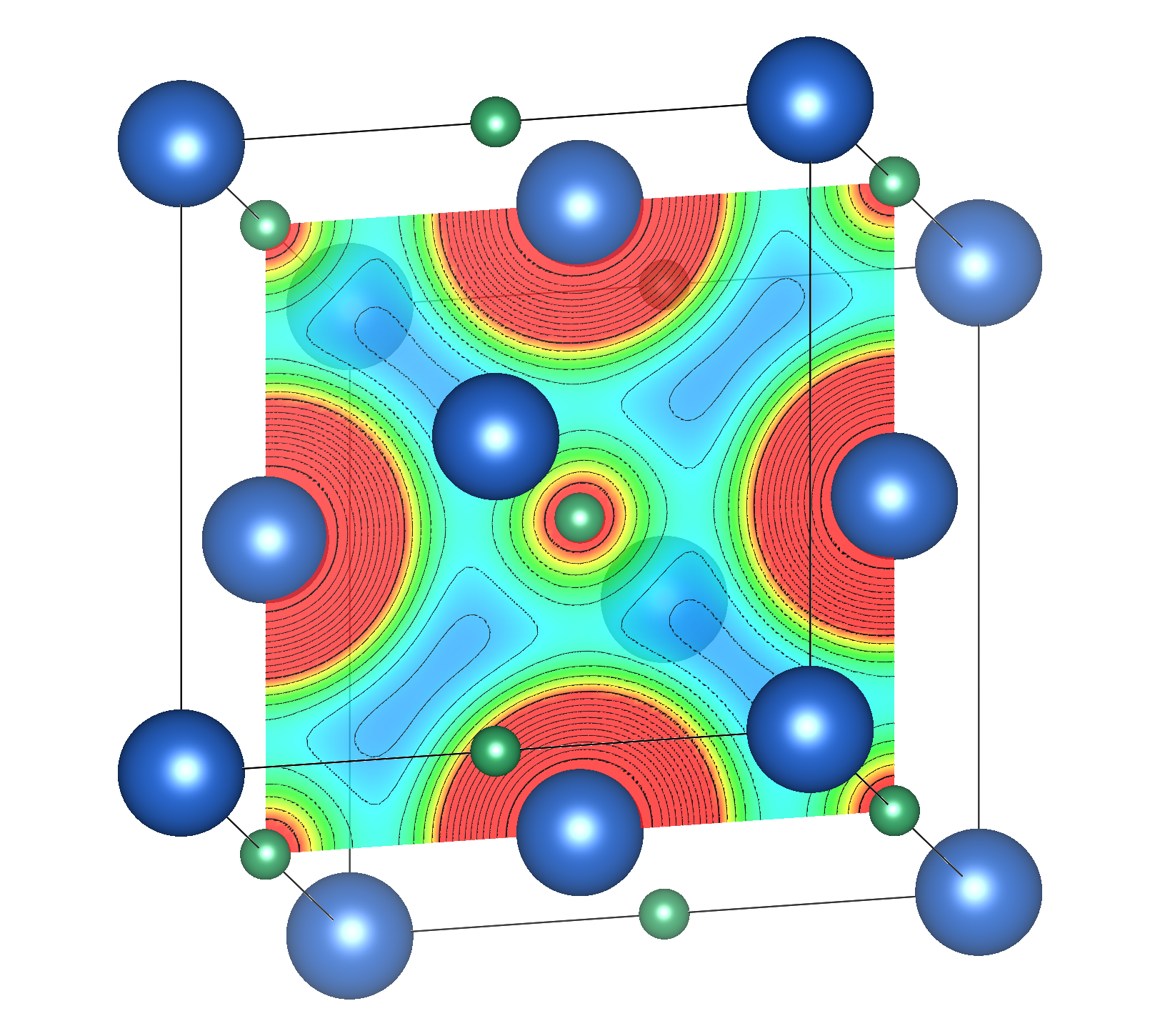}
    \caption{PdH$_{0.75}$}
\end{subfigure}
 \begin{subfigure}[t]{0.32\linewidth}
    \centering
    \includegraphics[width=1.0\linewidth]{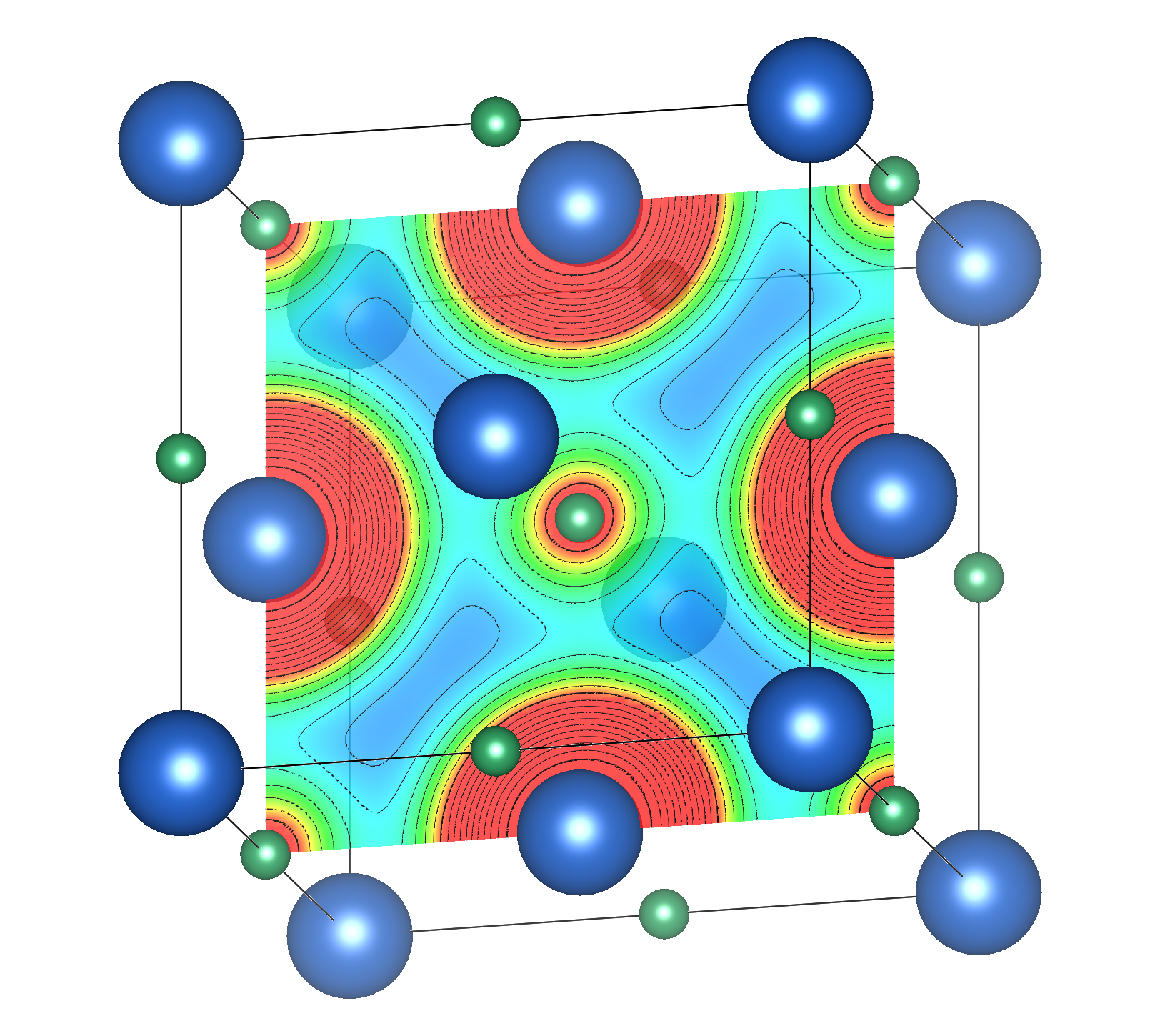}
    \caption{PdH}
\end{subfigure}
    \caption{Depiction of the calculated electronic densities for (a) Pd and (b)-(e) PdH$_x$ for different $x$, with H occupying the octahedral positions. Pd (H) atoms are depicted in blue (green).}
    \label{densities}
\end{figure*}

\begin{figure}[ht!]
    \centering
    \includegraphics[width=0.7\textwidth]{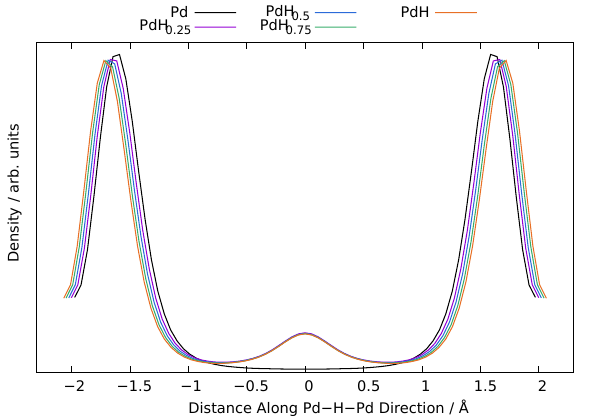}
    \caption{Plot of the calculated electronic densities along the Pd-H-Pd direction, i.e.\ (0.0,0.5,0.5) to (1.0,0.5,0.5), for Pd and PdH$_x$ for different $x$, with H occupying the octahedral positions. For ease of reading, the $x$ axis origin is at (0.5,0.5,0.5), i.e.\ centred around the H position where a H atom exists.}
    \label{tet}
\end{figure}

\bibliography{refs} 
\bibliographystyle{apsrev4-1}